\documentclass[preprint,nofootinbib,amsmath,amssymb,aps,prd]{revtex4}
\usepackage{braket}             % Allows Dirac notation
\usepackage{graphicx}           % Include figure files
\usepackage{bm}             % bold math
\usepackage[usenames, dvipsnames]{color}
\usepackage{hyperref}           % add hypertext capabilities
\usepackage{mathtools}
\usepackage[normalem]{ulem} %strikethrough text
\usepackage[caption=false]{subfig}
\usepackage{float}
\usepackage{graphicx}

\usepackage{soul}              % for crossing things out
\usepackage{colonequals} %Get a nice looking :=

\usepackage[usenames, dvipsnames]{xcolor}

\definecolor{capri}{rgb}{0.0, 0.75, 1.0}

\begin{document}

\title{Entanglement harvesting  of accelerated detectors versus  static ones in a thermal bath}
\author{Zhihong Liu$^{1}$, Jialin Zhang$^{1,2}$~\footnote{Corresponding author. jialinzhang@hunnu.edu.cn}and Hongwei Yu$^{1,2}$~\footnote{Corresponding author. hwyu@hunnu.edu.cn}}
\affiliation{$^1$ Department of Physics and Synergetic Innovation Center for Quantum Effects and Applications, Hunan Normal University, 36 Lushan Rd., Changsha, Hunan 410081, China\\
$^2$ Institute of Interdisciplinary Studies, Hunan Normal University, 36 Lushan Rd., Changsha, Hunan 410081, China}
\date{\today}
\begin{abstract}
We make a detailed comparison between entanglement
harvesting for uniformly accelerated detectors in vacuum  and static ones in a thermal bath at the Unruh
temperature and  find that,  for a small energy
gap relative to the  Heisenberg energy of the detectors, static detectors
in the thermal bath can harvest more entanglement and possess a
comparatively larger harvesting-achievable range than the uniformly
accelerated ones; however, as the  energy gap grows sufficiently
large, the uniformly accelerated detectors are instead likely to
harvest more entanglement and possess a relatively larger
harvesting-achievable range than inertial ones in the thermal bath. In comparison with static detectors in vacuum, there exist phenomena of acceleration-assisted entanglement harvesting  but never that of thermal-noise-assisted one.  A notably interesting feature is that, although both the amount of entanglement harvested and the harvesting-achievable interdetector separation for static detectors in a thermal bath are always a decreasing function of temperature,
they  are not  always so for uniformly accelerated detectors as acceleration (Unruh temperature) varies, suggesting the existence of the anti-Unruh effect in the entanglement harvesting phenomena of the accelerated detectors.
\end{abstract}

\maketitle

%========================================
%========================================

\section{Introduction}

 The vacuum state of any free quantum
field is entangled in the sense that  it can maximally violate
Bell's inequalities~\cite{Summers:1987}. It is known that the
Minkowski vacuum entanglement can be extracted by two particle
detectors via local interactions with vacuum quantum fields for a
finite time, even if the detectors are spacelike
separated~\cite{VAN:1991,Reznik:2003,BLHu:2012}. This phenomenon has
been dubbed entanglement
harvesting~\cite{Salton-Man:2015,Pozas-Kerstjens:2015}. Up to now,
the entanglement harvesting phenomenon has been extensively studied
by using the Unruh-DeWitt (UDW)  detector model in various circumstances, and
it is found that entanglement harvesting is quite sensitive to spacetime
dimensionality~\cite{Pozas-Kerstjens:2015},
topology~\cite{EDU:2016-1} and
curvature~\cite{Steeg:2009,Nambu:2013,Kukita:2017,Ng:2018,Ng:2018-2,Zhjl:2018,Zhjl:2019,Robbins:2020jca,Gallock-Yoshimura:2021},
 the presence of boundaries~\cite{CW:2019,CW:2020,Zhjl:2021},  and the characters of detectors such as
their intricate motion~\cite{Salton-Man:2015,Zhjl:2020,Zhjl:2021,Zhjl:2022,R.B.Mann:2022} and
 energy gap~\cite{Zhjl:2022.4,Maeso:2022}.

For the influence of  intricate motion of detectors on entanglement harvesting, let us note that the range finding of entanglement for  detectors (the harvesting-achievable
range of interdetector separation)  in different
scenarios of uniform acceleration
has been discussed in Refs.~\cite{Salton-Man:2015} by using the
saddle-point approximation. It was argued
there that entanglement harvesting can be
enhanced only in a special acceleration scenario, i.e., the
antiparallel acceleration. More recently, a more comprehensive
investigation in the entanglement harvesting phenomenon which
includes not only  the  harvesting-achievable separation range but
also the amount of entanglement harvested has been  performed with
both analytical analyses and  numerical calculations in
Refs.~\cite{Zhjl:2021,Zhjl:2022}. It is demonstrated  that
acceleration increases the amount of harvested entanglement and
enlarges the harvesting-achievable range in all the scenarios (parallel, antiparallel and mutually perpendicular
acceleration) once the energy gap is sufficiently large relative
to the interaction duration~\cite{Zhjl:2022}, and there is
no evidence of the phenomena of entanglement resonance argued
previously using the saddle-point
approximation~\cite{Salton-Man:2015}, which seems to suggest that
the conclusions based upon such an approximation may not be
reliable.

It is worth noting that, associated with acceleration, there is a striking effect  in quantum field theory, known as the Unruh effect,
 which predicts that
a uniformly accelerated observer in the Minkowski vacuum will
perceive a thermal bath of particles at a temperature  proportional to its proper
acceleration (the Unruh
temperature)~\cite{Unruh:1976}.  The Unruh effect establishes an amazing equivalence  between accelerated motion and  a thermal bath in terms of the response rate of an Unruh-Dewitt particle detector. Then, an interesting question
naturally  arises as to whether the entanglement harvesting
phenomena that involve two uniformly accelerated detectors would
also show such an equivalence of acceleration and thermal bath.

In the present paper, we will try to answer the  question by making
a detailed comparison between entanglement harvesting for
uniformly accelerated detectors in the Minkowski vacuum and static
ones in a  thermal bath in Minkowski spacetime, focusing upon  the
amount of entanglement harvested and the harvesting-achievable range
of interdetector separation. Let us note that  the
harvesting-achievable separation range of two static detectors in a
Minkowski thermal bath has been studied in Refs.~\cite{Steeg:2009,
Salton-Man:2015} by using certain method  of approximation, such as
the Taylor expansion or the saddle-point approximation, and it is
implied that  the static detectors in the Minkowski thermal bath at
the Unruh temperature always have a comparatively larger
harvesting-achievable range than the (parallel) accelerated ones~\cite{
Salton-Man:2015}. This conclusion, as we will demonstrate later, is, however, not universal but rather crucially dependent on the energy gap of the detectors.

In contrast, we  will approach the problem in the present paper with a more reliable
strategy. Specifically, our  investigation will be carried out with
numerical integration rather than  the saddle-point approximation adopted in Ref.~\cite{Salton-Man:2015} or the Taylor
expansion in Refs.~\cite{Steeg:2009}. Moreover, the amount of
entanglement harvested will  also be  studied and cross-compared.
We will demonstrate  that the entanglement
harvesting of accelerated detectors in terms of both the harvested
entanglement amount and the harvesting-achievable separation range
displays features  fundamentally  different  from those  of static
detectors in a thermal bath at the Unruh temperature.  The energy
gap of the detectors plays a critical role in determining which
scenario for the detectors is more conducive to entanglement
harvesting.  Our results suggest that  equivalence between
acceleration and thermal bath as seen by a single detector via the
detector's response  is lost when the entanglement harvesting of two
detectors are considered. Let us note here  that the loss  of the equivalence between temperature and acceleration  in the entanglement  dynamics for two detectors  has been discussed within the framework of open quantum systems in Refs.~\cite{Zhjl:2007,Hu:2015,Lima:2020}.

It is now worthwhile to point out that  there is a circular  version of the Unruh effect for detectors undergoing circular motion with constant centripetal acceleration instead of linear motion with uniform acceleration, which predicts that the detectors will also perceive  radiation, albeit with a non-Planckian spectrum~\cite{Bell:1983,Biermann:2020}, and it has been argued that  if one takes a circularly orbiting electron in a constant external magnetic field as the Unruh-DeWitt detector then the Unruh effect physically coincides with the  Sokolov-Ternov effect~\cite{Akhmedov:2007,Akhmedov:2008}.  Let us also note that the entanglement harvesting  phenomenon  for circularly moving detectors has recently been studied in Ref.~\cite{Zhjl:2020}, where it was found  that  centripetal acceleration has significant impacts on entanglement harvesting and
 the amount of harvested entanglement rapidly degrades  with increasing acceleration or interdetector separation.

The rest of the paper is organized as follows. We briefly review, in Sec. II,  the basics of the UDW detector
model and the protocol of entanglement harvesting. In Sec. III,
we, respectively, calculate and compare the entanglement harvested by uniformly
accelerated detectors in the Minkoswki vacuum and static
ones in a thermal bath at the Unruh temperature.
Approximate analytical results  will be presented in some particular
cases along with  detailed numerical calculations
for the two scenarios in general. We end with a conclusion  in
Sec. IV. For convenience, we adopt  the natural units
$\hbar=c=k_B=1$ throughout this paper.

%%%%%%%%%%%%%%%%%%%%%%%%%%
\section{Basic formalism}
We consider two identical two-level Unruh-DeWitt detectors $A$ and
$B$ interacting locally with a massless scalar field
$\phi[x_{D}(\tau)]$. The spacetime trajectory of the
detector, $x_{D}(\tau)$ ($D\in\{A,B\}$), is parametrized by
its proper time $\tau$. Then, the interaction Hamiltonian of the
detector and the field is given by
\begin{equation}\label{Int1}
 H_{D}(\tau)=\lambda \chi(\tau)\left[e^{i \Omega\tau} \sigma^{+}+e^{-i \Omega\tau} \sigma^{-}\right] \phi\left[x_{D}(\tau)\right]\;,~~ D\in\{A,B\},
 \end{equation}
where $\lambda$ is the coupling strength, $\chi(\tau)=\exp
[-{\tau^{2}}/(2\sigma^{2})]$  is the Gaussian switching function with
parameter $\sigma$  controlling  the duration of the interaction,
and $\Omega$ is the detectors' energy gap between the ground state
$|0_{D}\rangle$ and the excited state $|1_{D}\rangle$
($D\in\{A,B\}$). Here,
$\sigma^{+}=|1_{D}\rangle\langle0_{D}|$ and
$\sigma^{-}=|0_{D}\rangle\langle1_{D}|$ denote SU(2) ladder
operators.

Suppose that initially the two detectors are prepared in their
ground state.
Then, to leading order in the coupling strength, the density matrix for the final state of the detectors  can be obtained by tracing
out the field degrees of freedom in the
basis$\{\ket{0_A}\ket{0_B},\ket{0_A}\ket{1_B},\ket{1_A}\ket{0_B},\ket{1_A}\ket{1_B}\}$
\cite{Pozas-Kerstjens:2015,Zhjl:2019},
\begin{align}\label{rhoAB}
\rho_{AB}=\begin{pmatrix}
1-P_A-P_B & 0 & 0 & X \\
0 & P_B & C & 0 \\
0 & C^* & P_A & 0 \\
X^* & 0 & 0 & 0 \\
\end{pmatrix}+{\mathcal{O}}(\lambda^4)\;,
\end{align}
where the transition probability $ P_D$ reads
\begin{equation}\label{PAPB}
 P_D:=\lambda^{2}\iint d\tau d\tau' \chi(\tau) \chi(\tau') e^{-i \Omega(\tau-\tau')}
 W\left(x_D(t), x_D(t')\right)\quad\quad D\in\{A, B\}\;,
\end{equation}
and the quantities $C$ and $X$, which characterize correlations, are
given by
\begin{align}
C &:=\lambda^{2} \iint d \tau d \tau^{\prime} \chi(\tau) \chi(\tau')
e^{-i \Omega(\tau-\tau')} W\left(x_{A}(t), x_{B}(t')\right)\;,
\end{align}
\begin{align}\label{xxdef}
X:=-\lambda^{2} \iint d\tau d \tau' \chi(\tau)\chi(\tau')  e^{-i\Omega( \tau+\tau')}
\Big[\theta(t'-t)W\left(x_A(t),x_B(t')\right)+\theta(t-t')W\left(x_B(t'),x_A(t)\right)\Big]\;,
\end{align}
where $W(x,x')$ is the Wightman function of the quantum fields
[e.g., for the quantum field in the Minkowski vacuum state,
$W(x,x'):=\bra{0_M}\phi(x)\phi(x')\ket{0_M}$] and $\theta(t)$
represents the Heaviside theta function. Note that the detector's
coordinate time is a function of its proper time in  the above equations, i.e., $t=t(\tau)$.
The amount of harvested entanglement can be quantified  by
concurrence~\cite{WW}, which, for the density matrix~(\ref{rhoAB}),
is given by~\cite{EDU:2016-1,Zhjl:2018}
\begin{equation}\label{condf}
\mathcal{C}(\rho_{A B})=2 \max \Big[0,|X|-\sqrt{P_{A}
P_{B}}\Big]+\mathcal{O}(\lambda^{4})\;.
\end{equation}
 Obviously,  the amount of entanglement
acquired by the detectors is determined by the competition between
the correlation term $X$ and the geometric mean of
the transition probabilities, which in general depend on the motion
status and the energy gap of the detectors.

\section{Entanglement harvesting for  detectors in uniform acceleration and in a thermal bath}
In this section, we are going to analyze and compare entanglement harvesting for detectors in uniform acceleration and in a thermal bath. For this purpose, we first need to
calculate  $P_D$ and $X$ in  the expression of concurrence~(\ref{condf}).

\subsection{ Scenario of uniform acceleration}
We  assume  the two  identical detectors are accelerating along the
$x$ direction with acceleration $a$; then, the corresponding
trajectories  can be written as
\begin{align}\label{traj-aa1}
&x_A:=\{t=a^{-1}\sinh(a\tau_A)\;,~x=a^{-1}\cosh(a\tau_A)\;,~y=z=0\}\;,
\nonumber\\
&x_B:=\{t=a^{-1}\sinh(a\tau_B)\;,~x=a^{-1}\cosh(a\tau_B)+L\;,~y=z=0\}\;,
\end{align}
where $L$ represents the constant interdetector separation as measured in the
laboratory reference frame, i.e, that seen by an inertial observer.

The Wightman function for vacuum massless scalar fields in four-dimensional Minkowski spacetime reads~\cite{Birrell:1984}
\begin{align}\label{wigh-1}
W\left(x, x'\right)=&-\frac{1}{4 \pi^{2}}\frac{1}{(t-t'-i
\epsilon)^{2}-|\mathbf{x}-\mathbf{x'}|^{2}}\;.
\end{align}
Substituting  Eq.~(\ref{traj-aa1}) and  Eq.~(\ref{wigh-1})
into Eq.~(\ref{PAPB}), we obtain
~\cite{Zhjl:2020,Zhjl:2021}
\begin{align}\label{PD}
P_{D}=&\frac{\lambda^{2} T_{U} \sigma}{2\sqrt\pi} \int_{0}^{\infty}
d\tilde{s} \frac{\cos (\tilde{s} \gamma)e^{-\tilde{s}^{2}
\alpha}\left(\sinh ^{2}\tilde{s}-\tilde{s}^{2}\right)}{\tilde{s}^{2}
\sinh^{2} \tilde{s}}+\frac{\lambda^{2}}{4 \pi}\left[e^{-\Omega^{2}
\sigma^{2}}-\sqrt{\pi} \Omega \sigma \operatorname{Erfc}(\Omega
\sigma)\right] \;,
\end{align}
where
$T_{U}:=a/(2\pi)$ is the Unruh temperature,  $\gamma:=\Omega/(\pi T_{U})$, $\alpha:=1
/(2\pi{T_{U}}\sigma)^{2}$, and $\operatorname{Erfc}(z)$ denotes the
complementary error function. Similarly, the correlation term
 $X$, denoted here by $X_{acc}$  for acceleration, is given
 by~\cite{Zhjl:2022}
\begin{align}\label{Xacc}
X_{acc}&=-\frac{\lambda^2 T_{U}^2}{8}\int_{0}^{\infty}
d\tilde{y}\int_{-\infty}^{\infty}d\tilde{x}
e^{-\frac{\tilde{x}^2+\tilde{y}^2}{4\sigma^2}-i
\tilde{x}\Omega}F(\tilde{x},\tilde{y})\;,
\end{align}
where
\begin{align}\label{Facc}
F(\tilde{x},\tilde{y}):=&\Big\{\Big[\pi T_{U}L-e^{-\tilde{x}\pi
T_{U}}\sinh(\tilde{y}\pi T_{U})\Big]\Big[\pi T_{U}L+e^{\tilde{x}\pi
T_{U}}\sinh(\tilde{y}\pi T_{U})\Big]-i\epsilon\Big\}^{-1}
\nonumber\\
&+\Big\{\Big[\pi T_{U}L+e^{-\tilde{x}\pi T_{U}}\sinh(\tilde{y}\pi
T_{U})\Big]\Big[\pi T_{U}L-e^{\tilde{x} \pi T_{U}}\sinh(\tilde{y}\pi
T_{U})\Big]-i\epsilon\Big\}^{-1}\;.
\end{align}
Obviously, an  evaluation  of  Eqs.~(\ref{PD}) and (\ref{Xacc})
calls for numerical integration.

\subsection{Scenario of thermal bath}
The Wightman function for the   fields  in a thermal state
 at the Unruh temperature $T_{U}$ is given
by~\cite{Birrell:1984}
\begin{align}\label{wigh-0}
W\left(x, x'\right)=-\frac{1}{4
\pi^{2}}\sum_{m=-\infty}^\infty\frac{1}{(t-t'-im/T_{U}-i
\epsilon)^{2}-|\mathbf{x}-\mathbf{x'}|^{2}}\;.
\end{align}
For an inertial detector in a thermal bath, one
can verify  that the corresponding transition probability is exactly the
same as that of the detector in uniform acceleration, which is nothing but what
the well-known Unruh effect means~\cite{Birrell:1984}.
Adapted to the  two detectors at rest with an interdetector
separation $L$,  the Wightman function
Eq.~(\ref{wigh-0}) becomes
\begin{equation}\label{wigh-T2}
W\left(x,
x'\right)=\frac{T_U}{8\pi{L}}\Big\{\coth\Big[\pi{T_U}(L-t+t'+i\epsilon)\Big]+\coth\Big[\pi{T_U}(L+t-t'-i\epsilon)\Big]\Big\}\;.
\end{equation}
Inserting Eq.~(\ref{wigh-T2}) into  Eq.~(\ref{xxdef}), the
correlation term  $X$, denoted here by $X_{th}$ for a
 thermal bath at temperature $T_{U}$, can be written
as
\begin{align}
X_{th}=&-\frac{\lambda^2 e^{-\Omega ^2\sigma ^2}
T_{U}\sigma}{4\sqrt\pi L}\int_{0}^{\infty} ds
e^{-s^{2}/4\sigma^{2}}\left\{\coth\left[{\pi
T_{U}(L+s)}\right]+\frac{\cosh\left[{\pi
T_{U}(L-s)}\right]}{\sinh\left[\pi T_{U}(L-s)\right]-i
\epsilon}\right\} \;. \label{X-th}
\end{align}
 Equations~(\ref{Xacc}) and~(\ref{X-th}) show that
$|X_{acc}|$ is different  from $|X_{th}|$, and this  implies that
the entanglement harvesting in two scenarios will be not equivalent
in general.
For a qualitative understanding of the entanglement harvesting,
we first analytically estimate
$X_{acc}$ and $X_{th}$ in some special cases.

\subsection{ Analytical approximation}
For a small acceleration or  Unruh temperature  with respect to
the Heisenberg energy (i.e., $ a\ll1/\sigma$ or $T_{U}\ll1/\sigma$),
the correlation terms $X_{acc}$ and $X_{th}$ can be respectively
approximated as
\begin{align}\label{Xacc1}
X_{acc}\approx&-\frac{i\lambda^2\sigma e^{-(L^2+4\Omega^2\sigma^4)/(4\sigma^2)}
{\rm{Erfc}}[i{L}/{(2\sigma)}]}{4 L\sqrt{\pi}}-\frac{i\lambda^2 \pi^{3/2}{T_{U}}^2 }{24 L\sigma}e^{-(L^2+4\Omega^2\sigma^4)/(4\sigma^2)}\nonumber\\
&\times\Big\{\Big[3(4L^2\sigma^2+4\sigma^4-L^4)\Omega^2-9L^2-6\sigma^2\Big]\sigma^2+2L^4\Big\}
\;,
\end{align}
\begin{align}\label{Xth0}
X_{th}\approx-\frac{i\lambda^2\sigma
e^{-(L^2+4\Omega^2\sigma^4)/(4\sigma^2)}
{\rm{Erfc}}[i{L}/{(2\sigma)}]}{4 L\sqrt{\pi}}
-\frac{\lambda^2\pi{T_U}^2\sigma^2}{6}e^{-\Omega^{2}\sigma^2}\;.
\end{align}
If  a small interdetector separation ($L/\sigma\ll1$) is further assumed,
Eqs.~(\ref{Xacc1}) and~(\ref{Xth0}) can  then be simplified as
\begin{align}\label{Xacc11}
|X_{acc}|\approx&\frac{\lambda^2
e^{-\Omega^2\sigma^2}\sigma}{4\sqrt{\pi}L} -\frac{\lambda^2
e^{-\Omega^2\sigma^2}L}{16\pi^{3/2}\sigma}(\pi-2)+\frac{\lambda^2
e^{-\Omega^2\sigma^2}}{4
L}(2\Omega^2\sigma^2-1)\pi^{3/2}{T_U}^2\sigma^3 \;,
\end{align}
\begin{align}\label{Xth01}
|X_{th}|\approx&\frac{\lambda^2
e^{-\Omega^2\sigma^2}\sigma}{4\sqrt{\pi}L} -\frac{\lambda^2
e^{-\Omega^2\sigma^2}L}{16\pi^{3/2}\sigma}(\pi-2)
+\frac{\lambda^2e^{-\Omega^2\sigma^2}\pi^{1/2}\sigma }{6}L{T_U}^2
\;.
\end{align}
It is easy to find that
$|X_{acc}|-|X_{th}|\propto(2\Omega^2\sigma^2-1)$, suggesting that
$|X_{th}|$ is larger or smaller than $|X_{acc}|$  depending on
whether the energy gap $\Omega$ is smaller or larger than $1/(\sqrt
2\sigma)$, i.e., $\Omega\sigma<1/\sqrt 2$ or $\Omega\sigma>1/\sqrt
2$ . Let us recall the transition probability of the
detector~(\ref{PD}), which  can now be approximated as
\begin{align}\label{P0}
P_{D}\approx&\frac{\lambda^{2}}{4 \pi}\left[e^{-\Omega^{2}
\sigma^{2}}-\sqrt{\pi} \Omega \sigma \operatorname{Erfc}(\Omega
\sigma)\right]+\frac{\lambda^2\pi}{6}(T_{U}\sigma)^2e^{-\Omega^{2}\sigma^2}\;.
\end{align}
 Then, according to Eq.~(\ref{condf}), we can see  that the detectors both
in  uniform acceleration and in a thermal bath could
harvest entanglement when the acceleration (the Unruh temperature) and the
interdetector separation are small.  Furthermore,  static detectors  in a thermal bath can harvest more (less) entanglement than accelerated
detectors if $\Omega\sigma<1/\sqrt 2$  ($\Omega\sigma>1/\sqrt 2$).

On the other hand, if the acceleration (the Unruh temperature) and the
interdetector separation are large enough ($L/\sigma\gg
a\sigma>1$ and $a\gg\Omega$, to be precise), we have
\begin{align}\label{Xacc2}
|X_{acc}|\approx&\frac{\lambda^2 e^{-\Omega^2\sigma^2}\sigma^2}{2\pi
L^2}+\frac{\lambda^2\sigma^2}{2\pi^3{T_{U}}^2L^4}e^{(8\pi^2{T_{U}}^2-\Omega^2)\sigma^2}\cos(4\pi
T_{U}\Omega\sigma^2) \;,
\end{align}
\begin{align}\label{Xth02}
|X_{th}|\approx&\frac{\lambda^2T_{U}\sigma^2 }{2
L}e^{-\Omega^2\sigma^2} \;.
\end{align}
It is clear  that $|X_{acc}|$ and $|X_{th}|$ in general are
decreasing functions of  interdetector separation, and
$|X_{acc}|\propto{L}^{-2}$, $|X_{th}|\propto{L^{-1}}$. Therefore, as the
interdetector separation $L$ grows, $|X_{acc}|$ degrades more
rapidly than $|X_{th}|$. As a result,  the scenario of thermal bath
possesses a comparatively large harvesting-achievable
range of the interdetector separation as opposed to the scenario of uniform acceleration.

\subsection{Numerical results}
\begin{figure}[!htbp]
\centering \subfloat[\scriptsize$\Omega\sigma=0.50$, $a\sigma=2\pi
T_{U}\sigma=0.50$]
{\label{convsLL11}\includegraphics[width=0.32\linewidth]{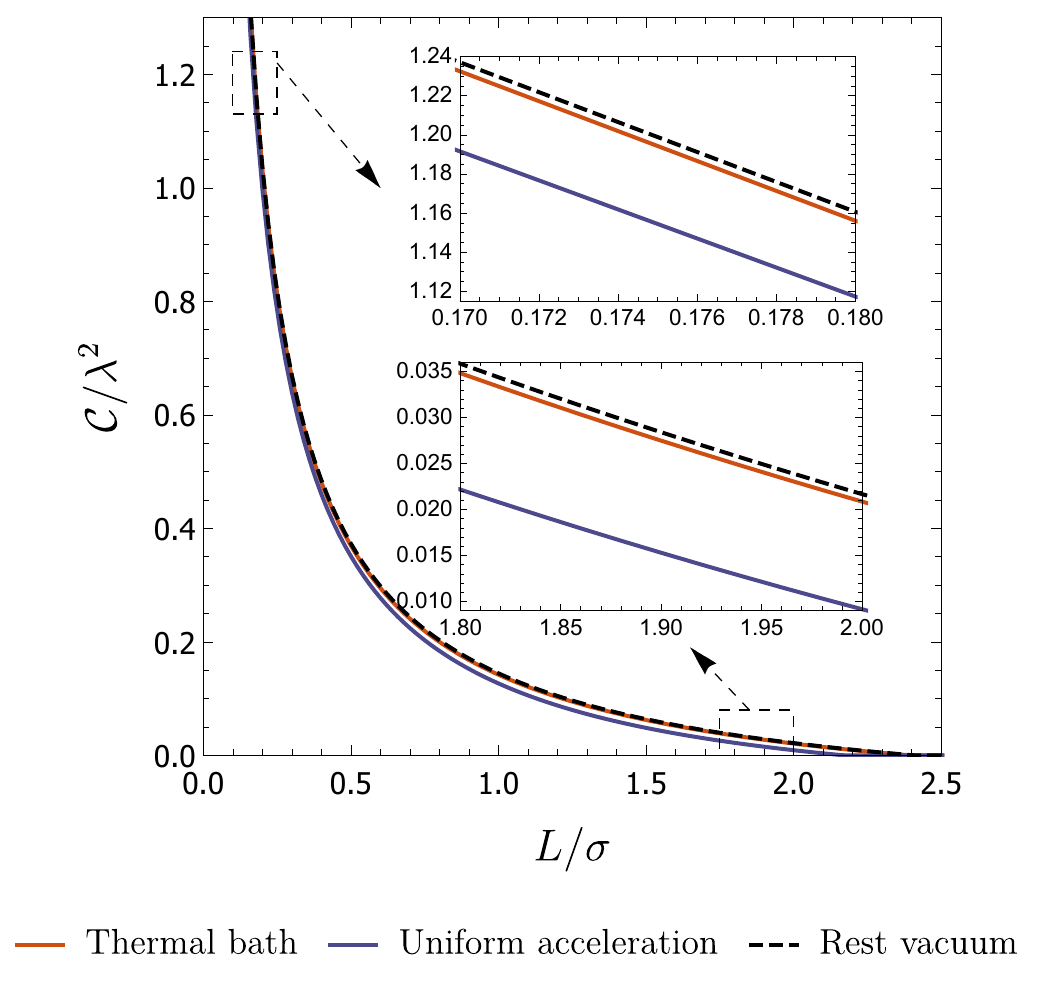}}\;
\subfloat[\scriptsize$\Omega\sigma=0.50$, $a\sigma=2\pi
T_{U}\sigma=1.00$]
 {\label{convsLL12}\includegraphics[width=0.32\linewidth]{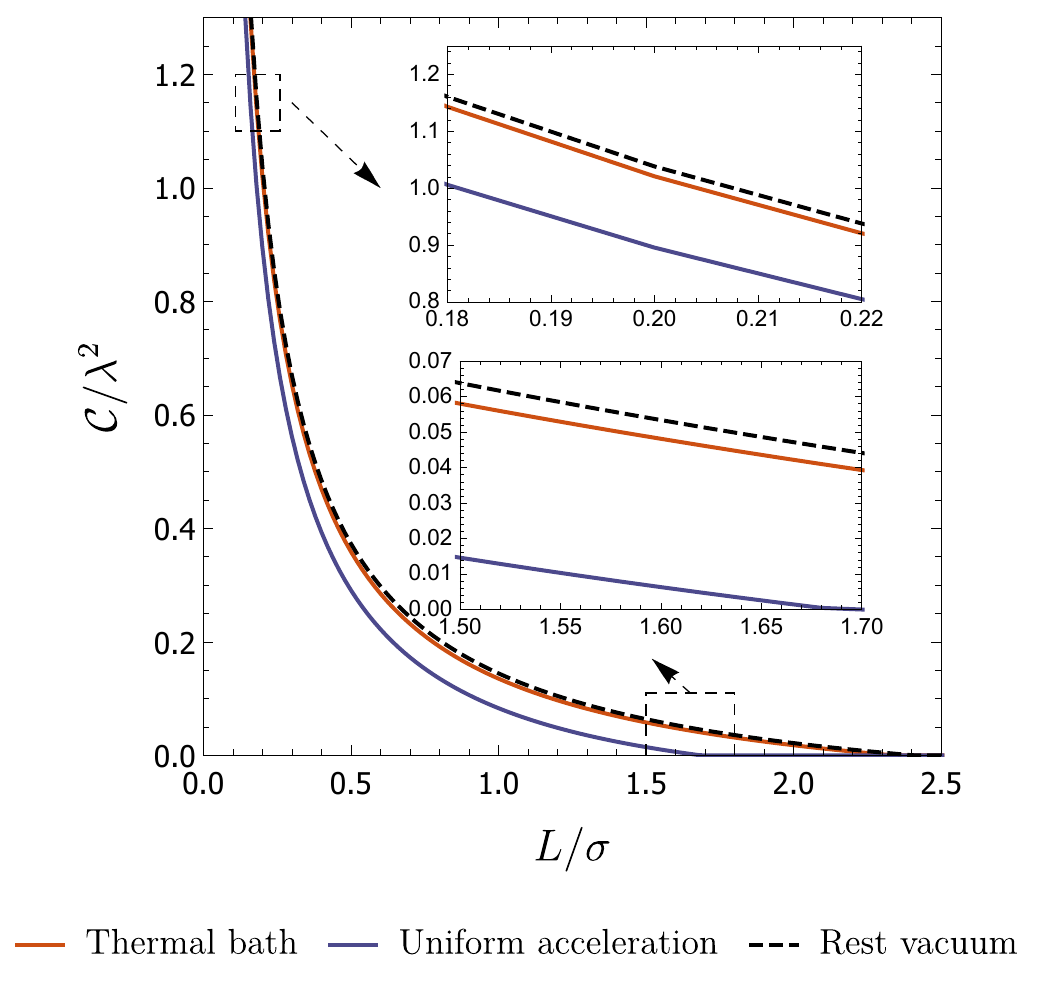}}\;
\subfloat[\scriptsize$\Omega\sigma=0.50$, $a\sigma=2\pi
T_{U}\sigma=3.00$]
 {\label{convsLL13}\includegraphics[width=0.32\linewidth]{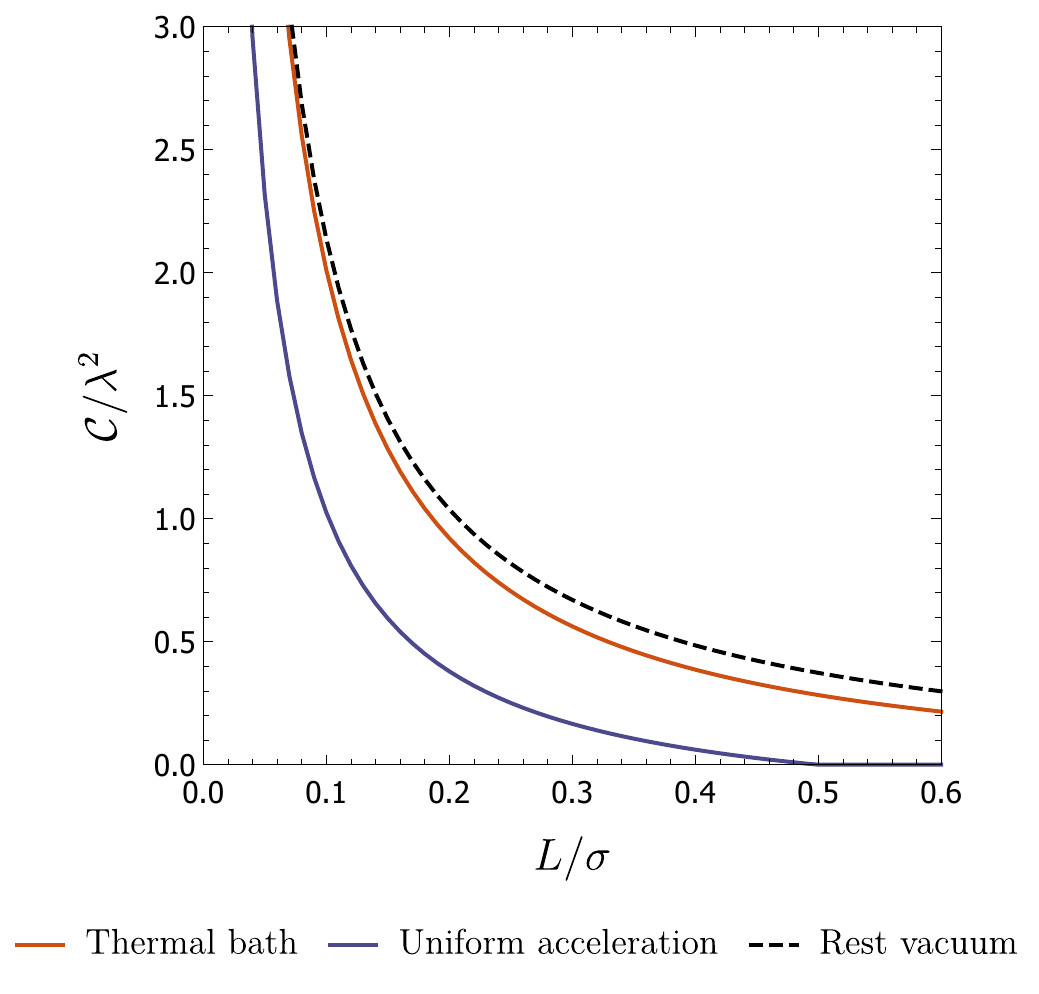}}
 \\
 \subfloat[\scriptsize$\Omega\sigma=1.20$, $a\sigma=2\pi T_{U}\sigma=0.50$]
 {\label{convsLL21}\includegraphics[width=0.32\linewidth]{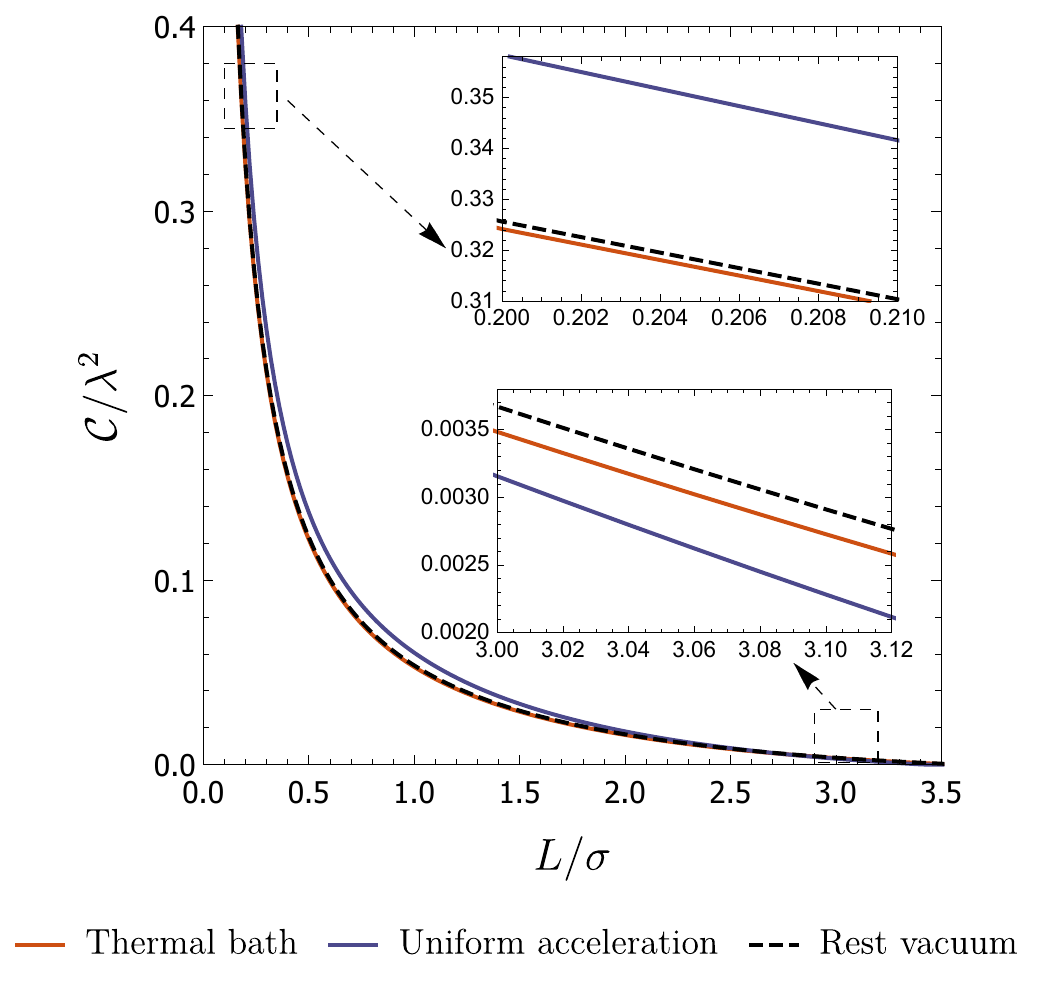}}\;
\subfloat[\scriptsize$\Omega\sigma=1.20$, $a\sigma=2\pi
T_{U}\sigma=1.00$]
{\label{convsLL22}\includegraphics[width=0.32\linewidth]{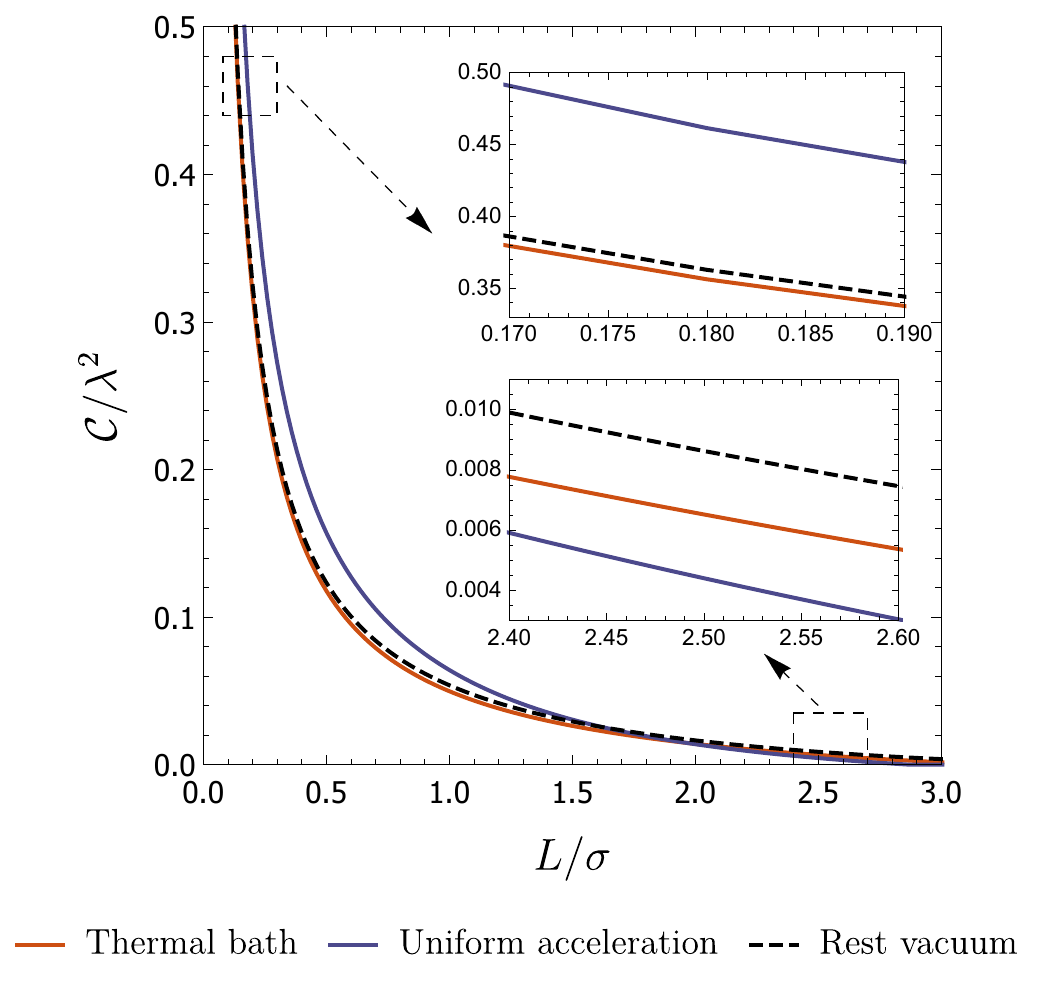}}\;
 \subfloat[\scriptsize$\Omega\sigma=1.20$, $a\sigma=2\pi T_{U}\sigma=3.00$]
 {\label{convsLL23}\includegraphics[width=0.32\linewidth]{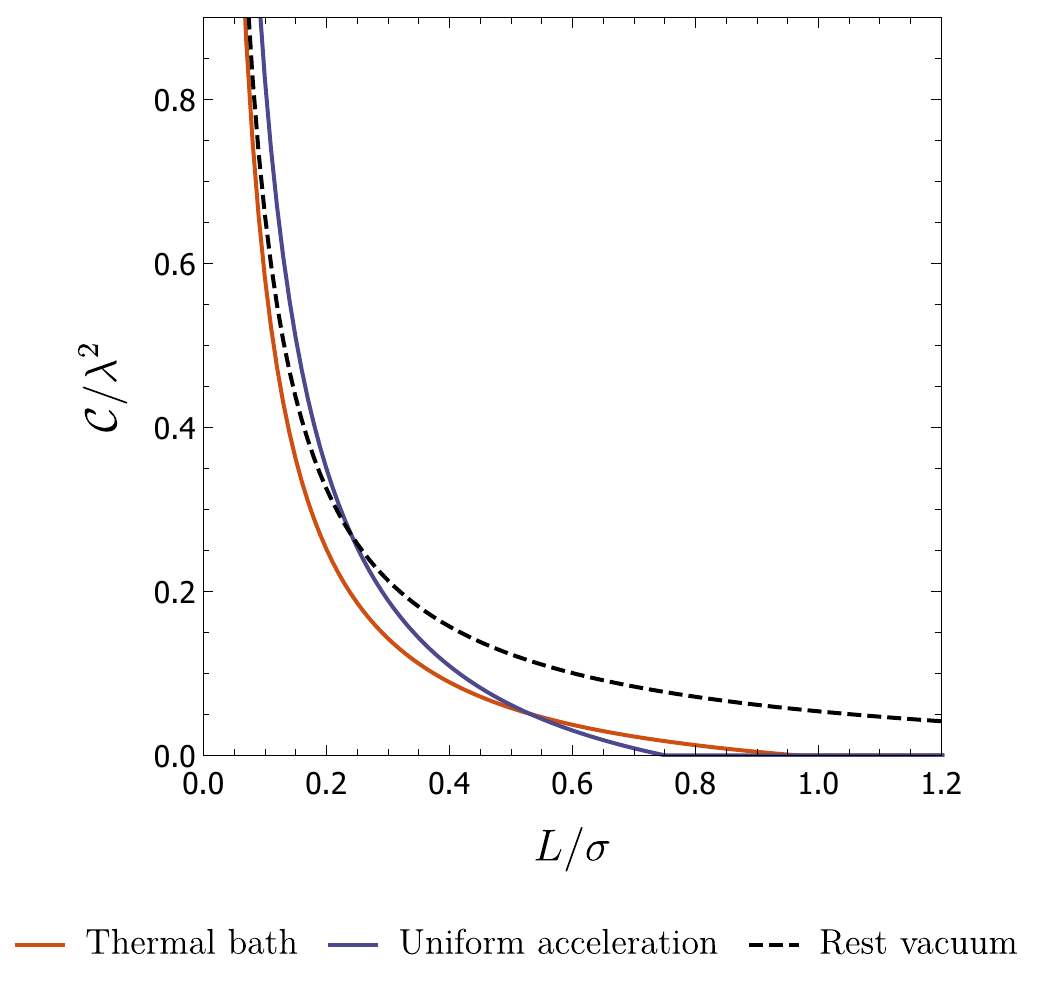}}
 \\
 \subfloat[\scriptsize$\Omega\sigma=2.00$, $a\sigma=2\pi T_{U}\sigma=0.50$]
 {\label{convsLL31}\includegraphics[width=0.32\linewidth]{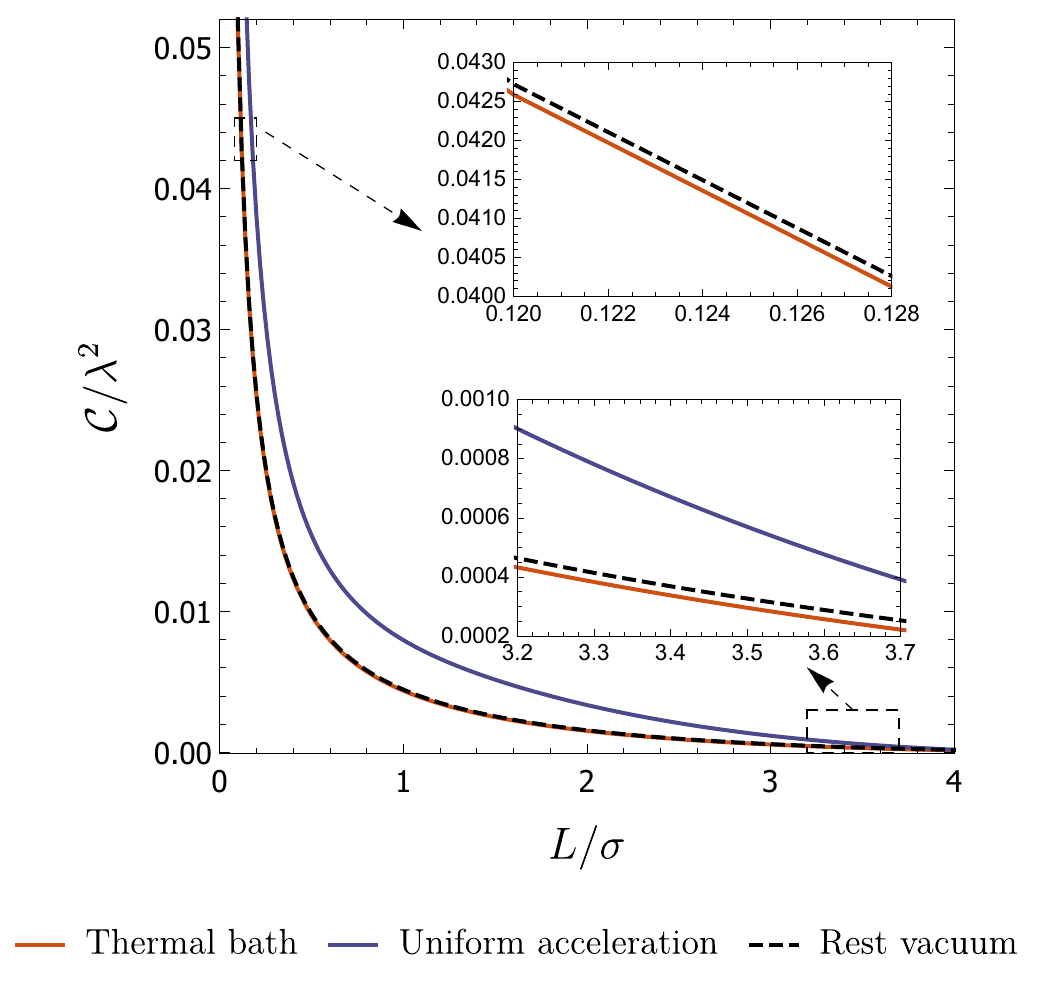}}\;
 \subfloat[\scriptsize$\Omega\sigma=2.00$, $a\sigma=2\pi T_{U}\sigma=1.00$]
  {\label{convsLL32}\includegraphics[width=0.32\linewidth]{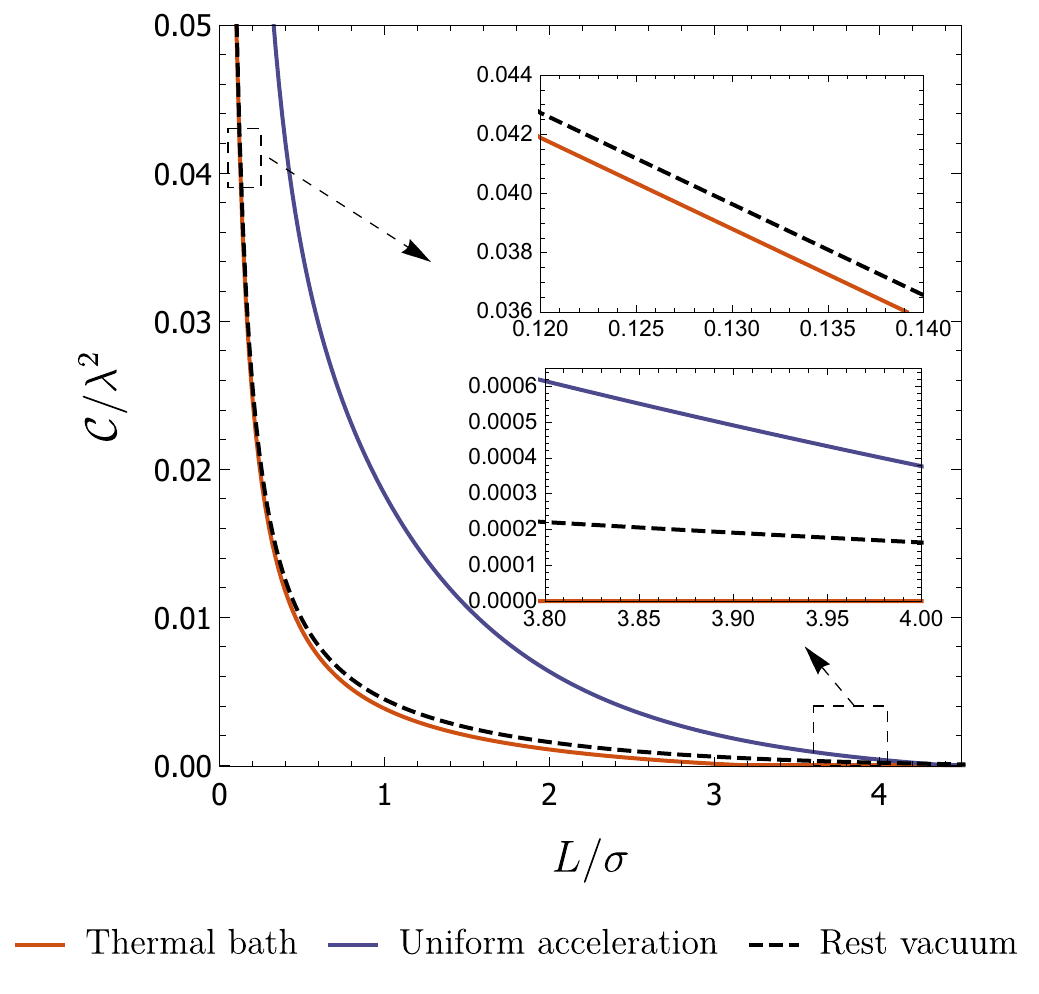}}\;
 \subfloat[\scriptsize$\Omega\sigma=2.00$, $a\sigma=2\pi T_{U}\sigma=3.00$]
 {\label{convsLL33}\includegraphics[width=0.32\linewidth]{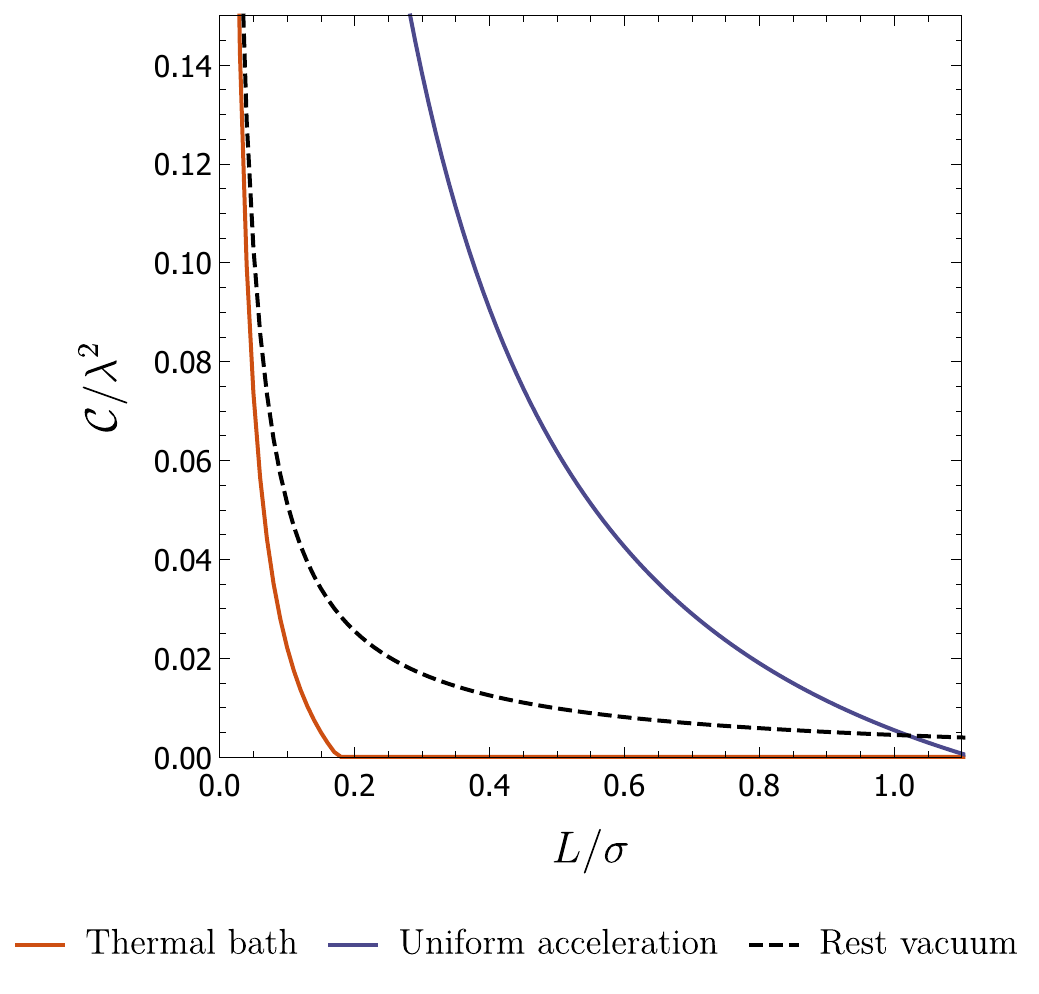}}
  \caption{The concurrence is plotted as a function of
the interdetector separation with $\Omega\sigma=\{0.50, 1.20, 2.00\}$
in top-to-bottom order and
$a\sigma=2\pi{T}_{U}\sigma=\{0.50,1.00,3.00\}$ in the left-to-right
order. Note that the dashed line corresponds to the scenario of detectors
at rest in the Minkowski vacuum. For convenience, all the physical
quantities are rescaled to be unitless in units of $\sigma$.
}\label{comp-d}
\end{figure}

\begin{figure}[!htbp]
\centering \subfloat[$\Omega\sigma=0.50$ and $L/\sigma=0.50$]
{\label{convsa11}\includegraphics[width=0.32\linewidth]{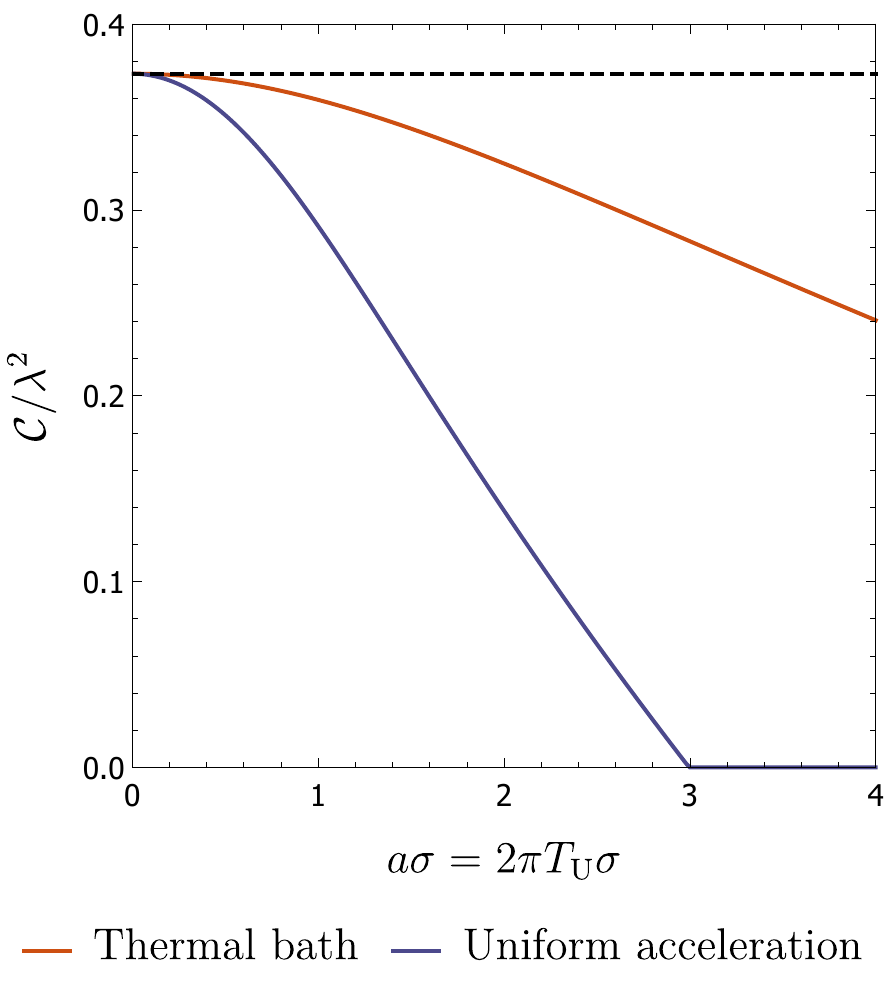}}\;
\subfloat[$\Omega\sigma=0.50$ and $L/\sigma=1.00$]
{\label{convsa12}\includegraphics[width=0.32\linewidth]{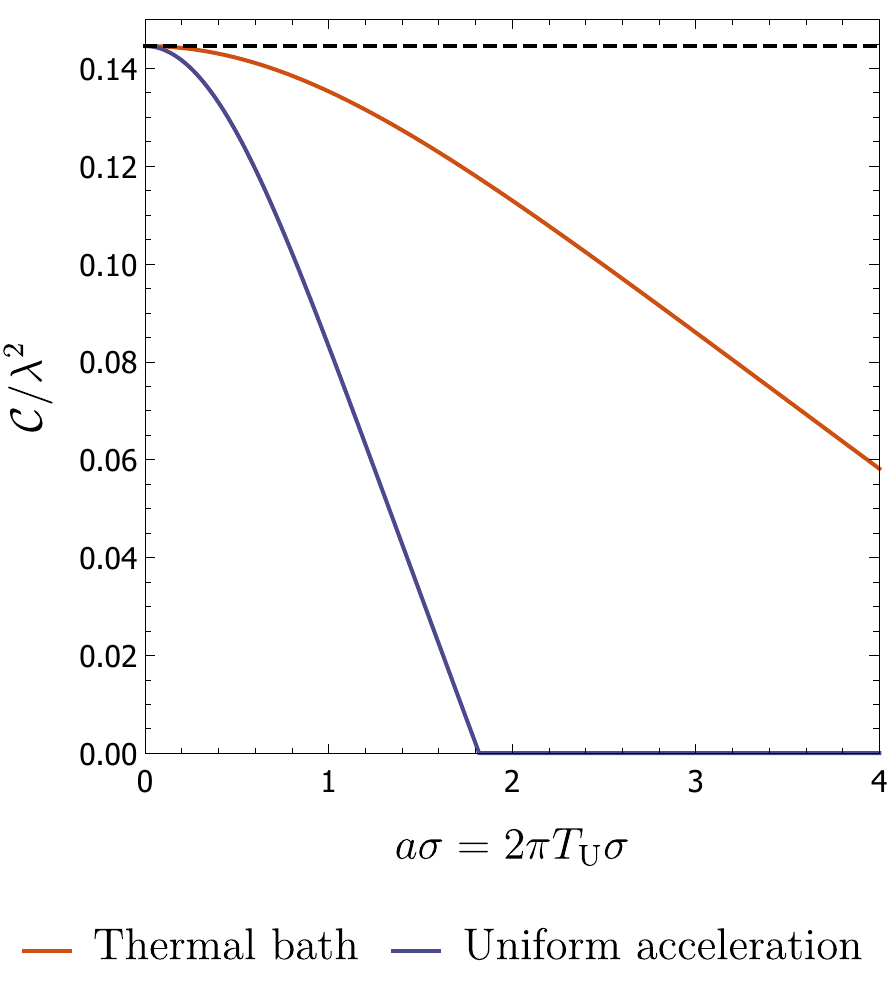}}\;
\subfloat[$\Omega\sigma=0.50$ and $L/\sigma=2.00$]
{\label{convsa13}\includegraphics[width=0.32\linewidth]{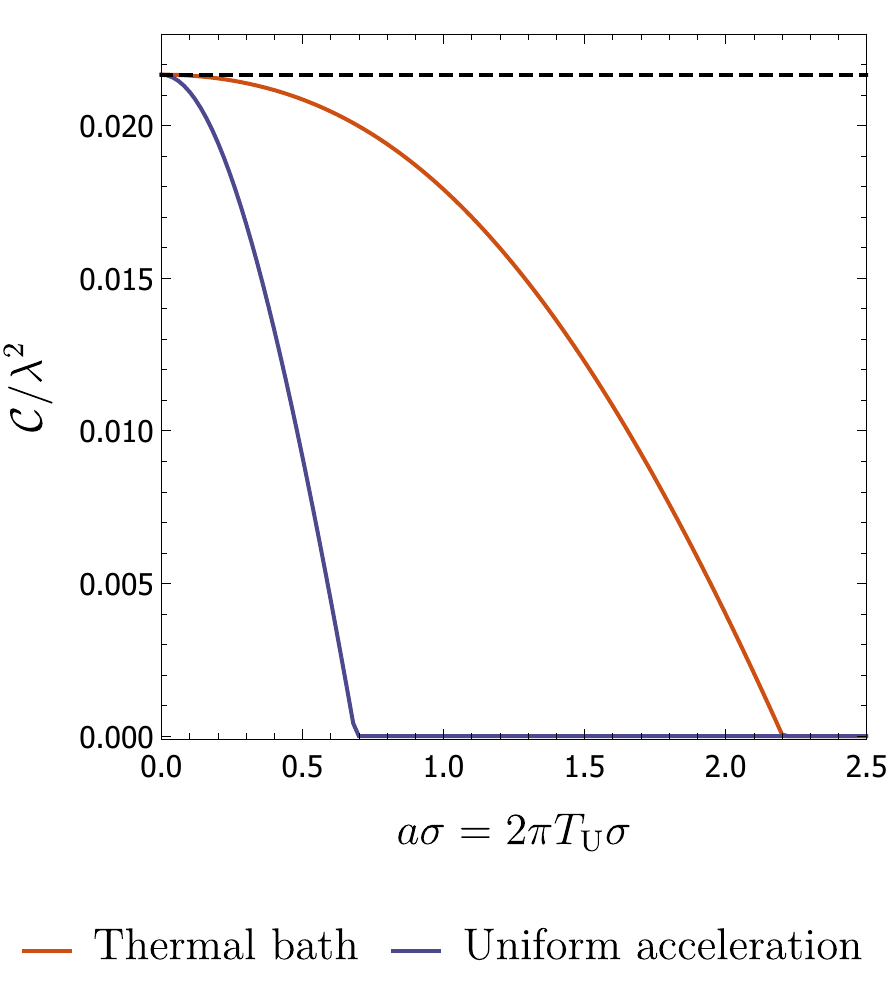}}
 \\
 \subfloat[$\Omega\sigma=1.20$ and $L/\sigma=0.50$]
 {\label{convsa21}\includegraphics[width=0.32\linewidth]{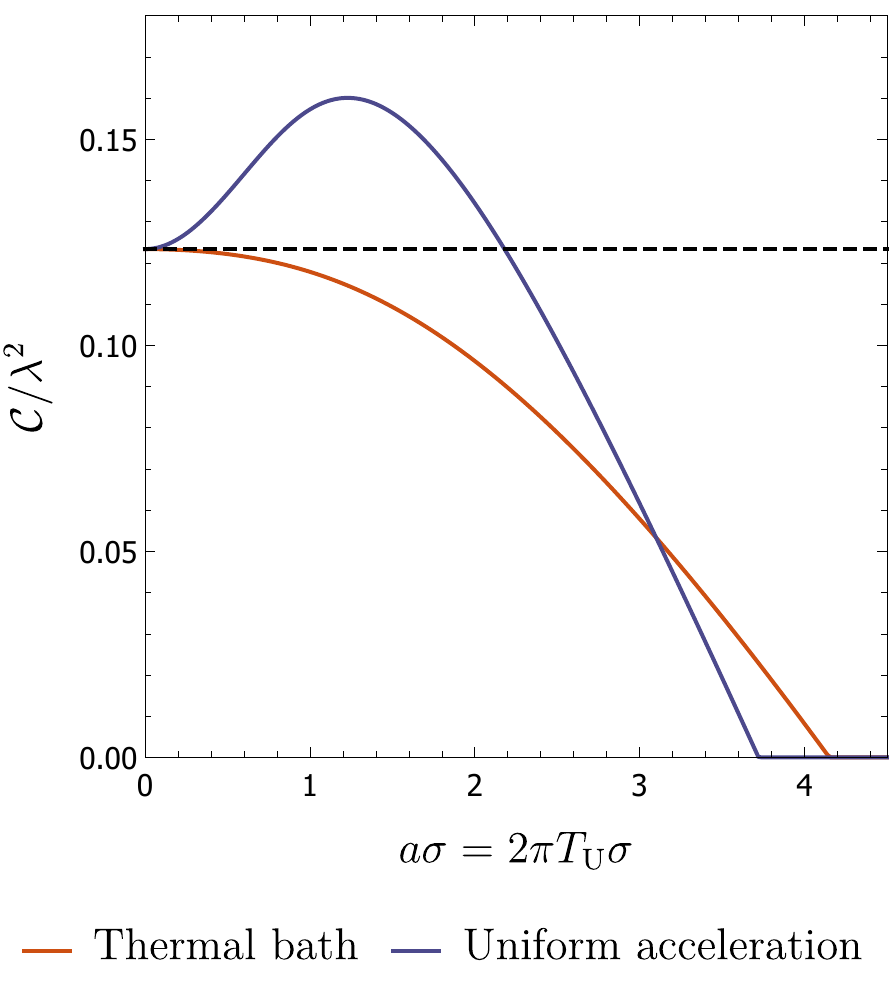}}\;
 \subfloat[$\Omega\sigma=1.20$ and $L/\sigma=1.00$]
 {\label{convsa22}\includegraphics[width=0.32\linewidth]{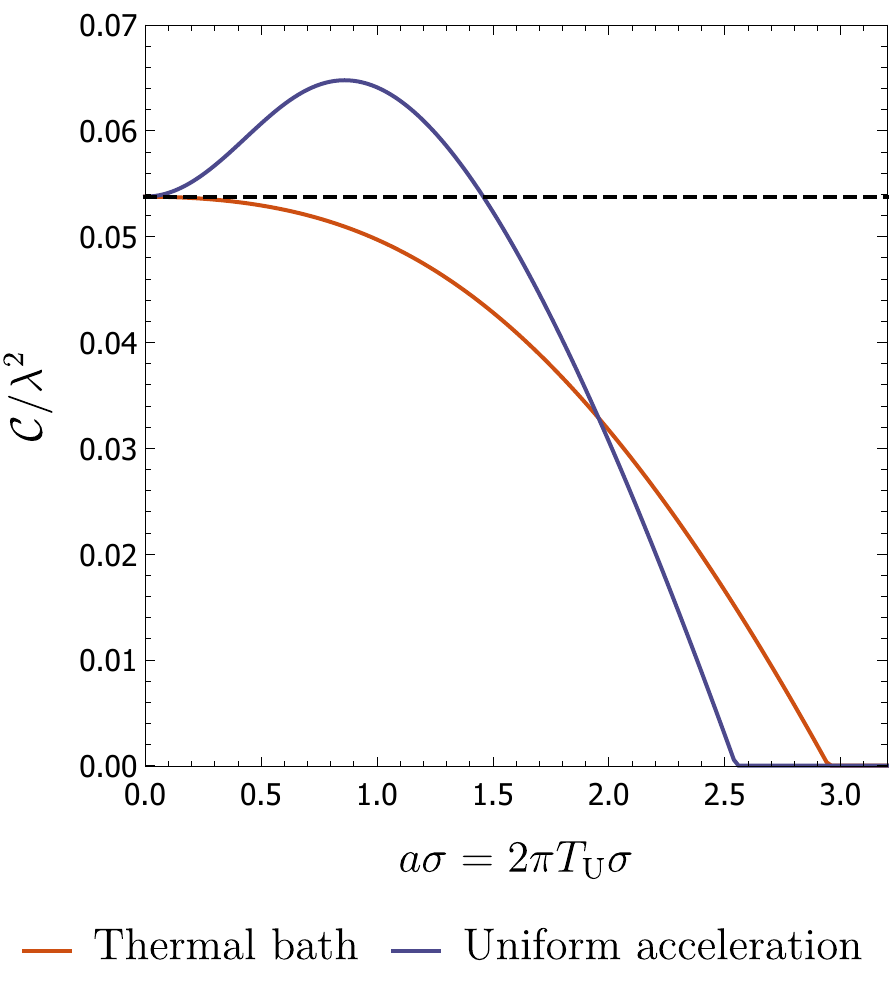}}\;
 \subfloat[$\Omega\sigma=1.20$ and $L/\sigma=2.00$] {\label{convsa23}\includegraphics[width=0.32\linewidth]{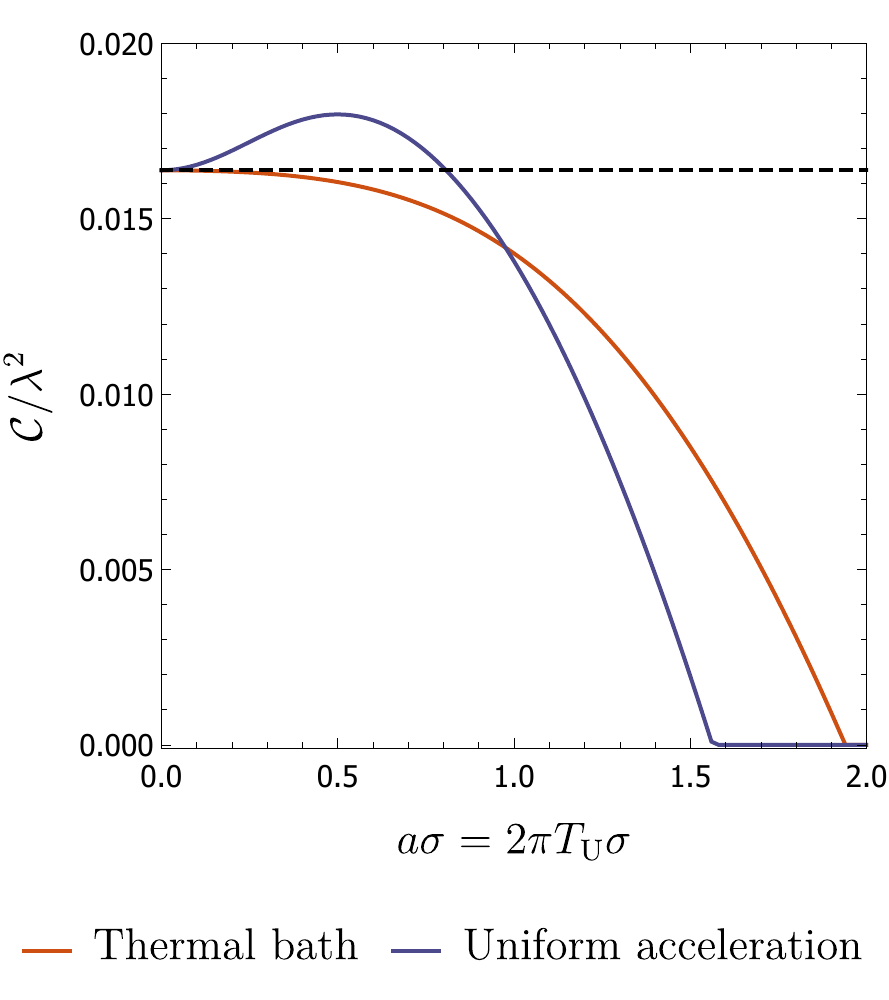}}
 \\
 \subfloat[$\Omega\sigma=2.00$ and $L/\sigma=0.50$]
 {\label{convsa31}\includegraphics[width=0.32\linewidth]{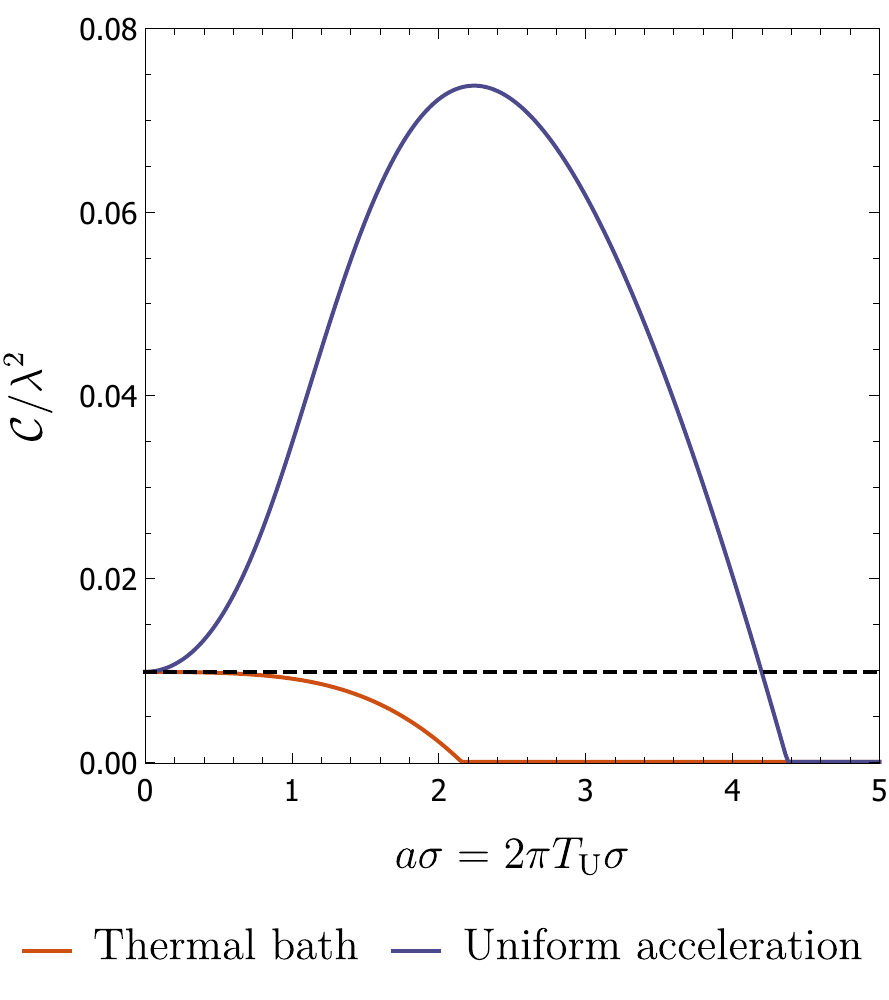}}\;
 \subfloat[$\Omega\sigma=2.00$ and $L/\sigma=1.00$]
 {\label{convsa32}\includegraphics[width=0.32\linewidth]{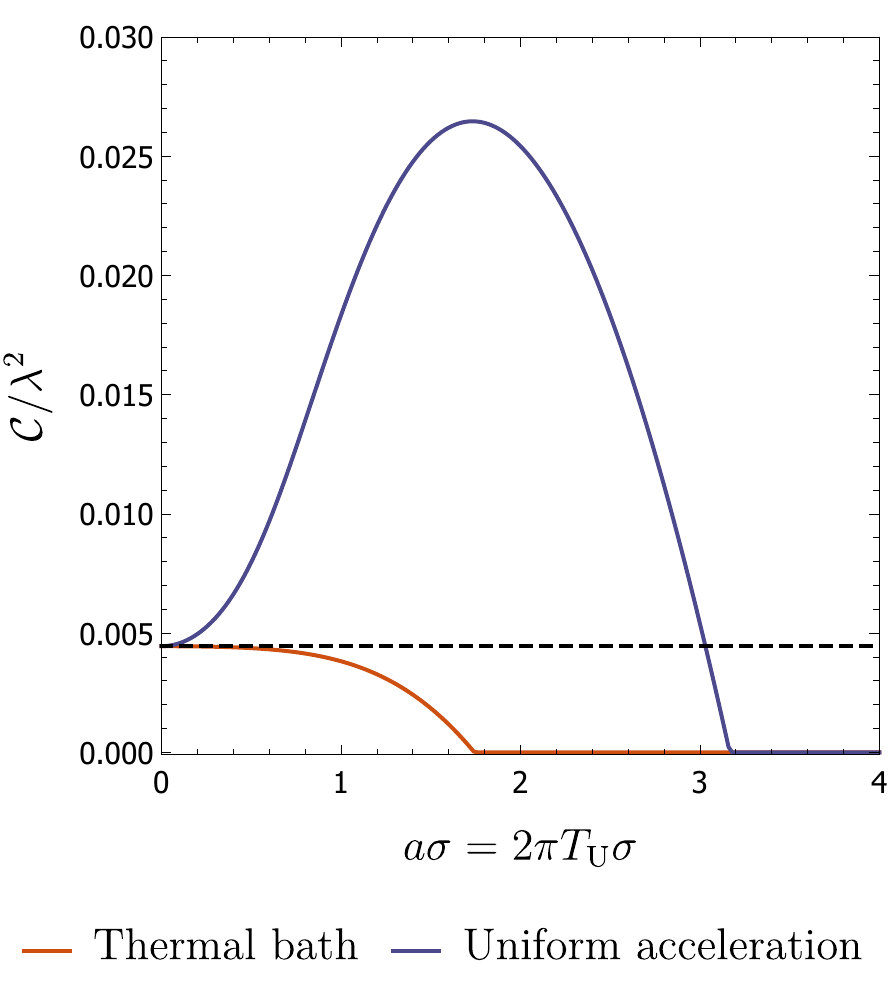}}\;
 \subfloat[$\Omega\sigma=2.00$ and $L/\sigma=2.00$] {\label{convsa33}\includegraphics[width=0.32\linewidth]{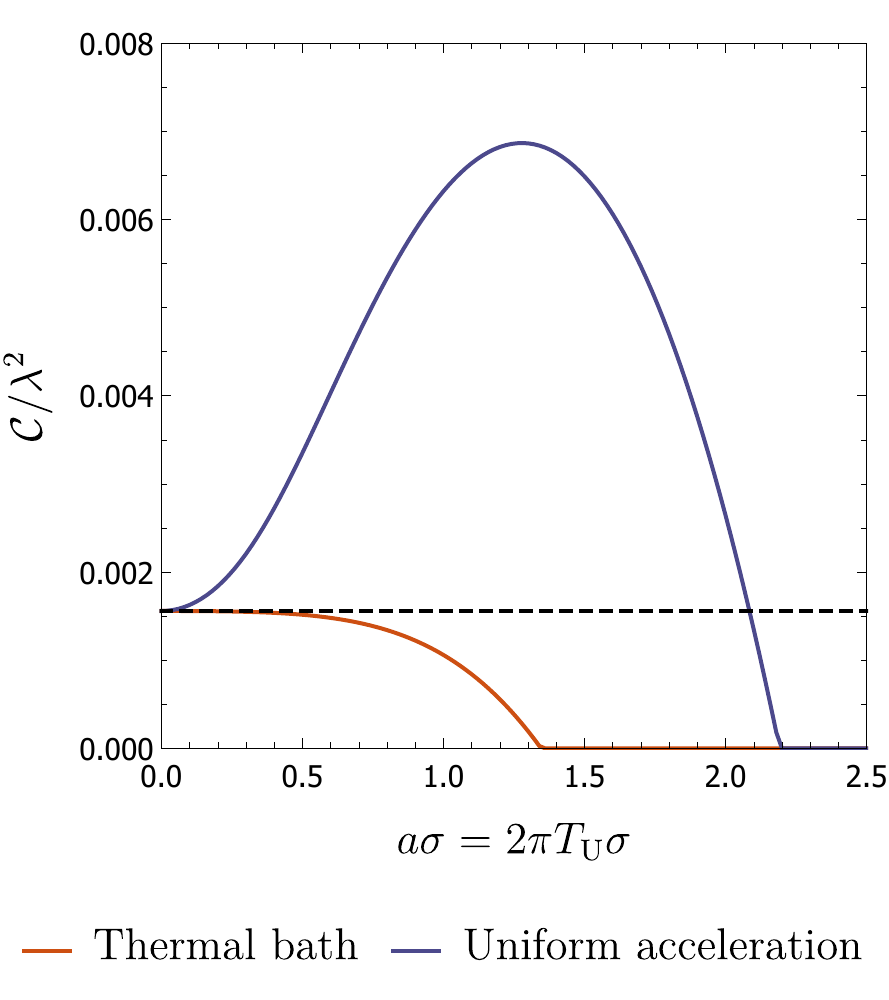}}
\caption{The plots of the concurrence versus the acceleration (or the Unruh temperature) with $\Omega\sigma=\{0.50,1.20,2.00\}$
in the top-to-bottom order and $L/\sigma=\{0.50,1.00,2.00\}$ in the
left-to-right order. The dashed lines in all plots indicate the case of detectors at rest ($a=0$) in Minkowski vacuum.
}\label{comp-A1}
\end{figure}

In general, the concurrence (\ref{condf}) can be
 obtained by numerical integration. As shown in
Figs.~\ref{comp-d} and~\ref{comp-A1}, the concurrence is
plotted as a function of  interdetector separation and
acceleration (the Unruh temperature), respectively. Obviously, the
concurrence is a monotonically decreasing function of
interdetector separation, which is in accordance with the
conclusions in the  previous
literature~\cite{EDU:2016-1,Zhjl:2021,Zhjl:2022}. The static
detectors  in a thermal bath always harvest less entanglement than
those in the Minkowski vacuum.  In other words, no thermal
noise-assisted entanglement harvesting occurs. In contrast, there
clearly exists acceleration-assisted entanglement harvesting as long
as the energy gap is large enough [see, for example, Fig.~\ref{comp-d}(i) or Fig.~\ref{comp-A1}(i)].  Comparing the case
of thermal bath with that of uniform acceleration, we find that
the detectors at rest in the Minkowski thermal bath could harvest
more entanglement if the energy gap $\Omega$ is much less than the
detectors' Heisenberg energy $1/\sigma$ ($\Omega\sigma<1/\sqrt{2}$, to be more precise). However, as the energy gap grows large
enough ($\Omega\sigma\gg1$), the accelerated detectors may instead harvest
comparatively more entanglement [e.g., see Fig.~\ref{comp-d}(g) or
Fig.~\ref{comp-A1}(g)]. This is in accordance with our analytical
analysis.  An interesting hallmark worthy of being noted here is that the amount of entanglement harvested is always a monotonically decrease function of temperature for  static detectors in a thermal bath, but it is not always so for uniformly accelerated detectors. In fact, for a sufficiently large energy gap, the harvested entanglement may first increase then decrease as acceleration grows [see, for example, Fig.~\ref{comp-A1}(e)]. This kind of nonmonotonicity of the harvested entanglement as a function of acceleration can be regarded the anti-Unruh effect~\cite{WGB:2016,Garay:2016,Liu:2016,ZhouYb:2021} in terms of the amount of entanglement harvested.

\begin{figure}[!htbp]
\centering
\includegraphics[width=0.65\linewidth]{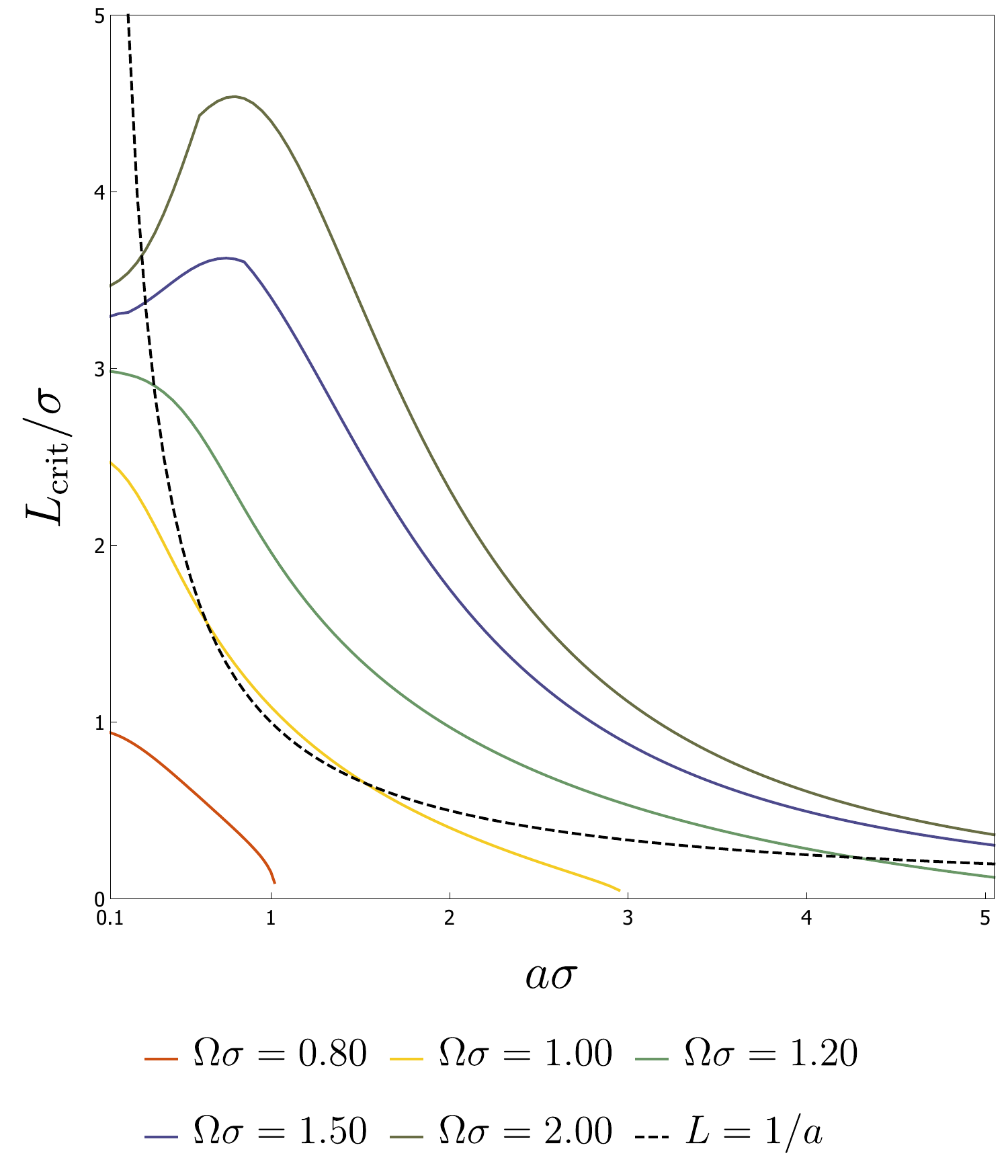}
\caption{The plots of $L_{crit}/\sigma$ versus the acceleration with
$\Omega\sigma=\{0.80,1.00,1.20,1.50,2.00\}$. The dashed black curve
represents the function $L=1/a$ that indicates the size of the
inertial region. Note that a non-negative $L_{crit}$, according to
the aforementioned discussion, should require
$\Omega\sigma>1/\sqrt{2}\sim 0.707$. }\label{L-d}
\end{figure}

We now turn our attention to the role of  interdetector separation in determining when the detectors in uniform
acceleration  could harvest more entanglement than static ones in
a thermal bath. For this purpose, we introduce
$L_{crit}$ to stand for the critical interdetector separation, below which (i.e., $L<L_{crit}$) the accelerated detectors could harvest more entanglement
than the inertial ones in a thermal bath.  As shown in
Fig.~\ref{L-d}, for not too large energy gap ($\Omega\sigma<1$),
$L_{crit}$  is a monotonically decreasing function of
acceleration (or the Unruh temperature), while as the energy gap
grows to large enough, although $L_{crit}$ on the whole still behaves as a decreasing
function of  acceleration, it  undergoes  some oscillation over
the regime of small $a\sigma$. Let us note that
when the interdetector separation $L$ is much smaller than the
characteristic length $1/a$ (i.e., $L\ll1/a$)
the interdetector interaction
between two accelerated detectors behaves
almost like that of two inertial detectors  in the Minkowski
vacuum~\cite{SCheng:2022}. Therefore, we could use
 $\sim 1/a$ as a characteristic separation to signal that the entanglement harvesting of two accelerated detectors would behave more or less like  that of two inertial detectors.
As long as the energy gap is not too small (i.e.,
$\Omega\sigma>1/\sqrt{2}$), Fig.~\ref{L-d} shows that when the
interdetector separation lies in the inertial regime ($L\ll1/a$),
the detectors in the acceleration scenario could harvest more
entanglement due to the fact that there does not exist thermal noise-assisted
entanglement harvesting but rather thermal noise-caused degradation. Meanwhile, the curve of $L_{crit}$ seems to have a
chance to be above the curve of $L=1/a$ if the energy gap is
large enough. This implies that a pair of accelerated detectors
separated by a noninertial distance could instead harvest
comparatively more entanglement due to the effect of
acceleration-assisted entanglement harvesting.

\begin{figure}[!htbp]
\centering
\subfloat[$\Omega\sigma=0.50$]{\label{Lmax-d11}\includegraphics[width=0.32\linewidth]{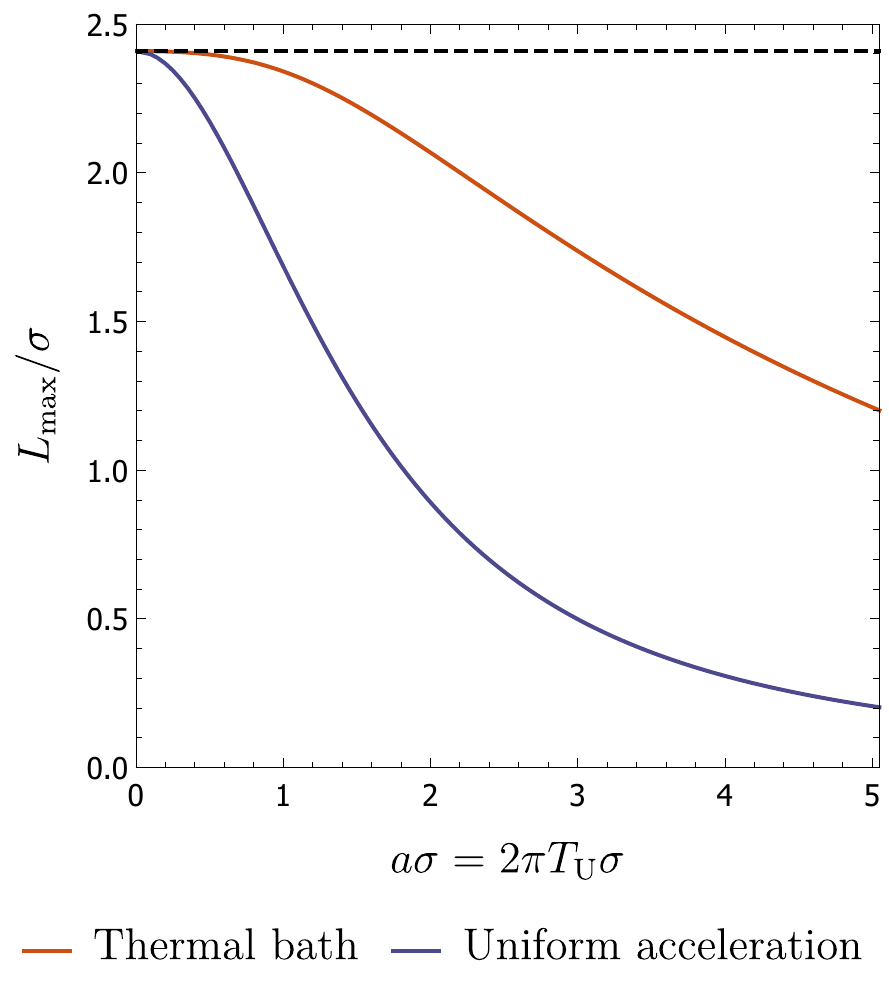}}\;
 \subfloat[$\Omega\sigma=1.00$]{\label{Lmax-d12}\includegraphics[width=0.32\linewidth]{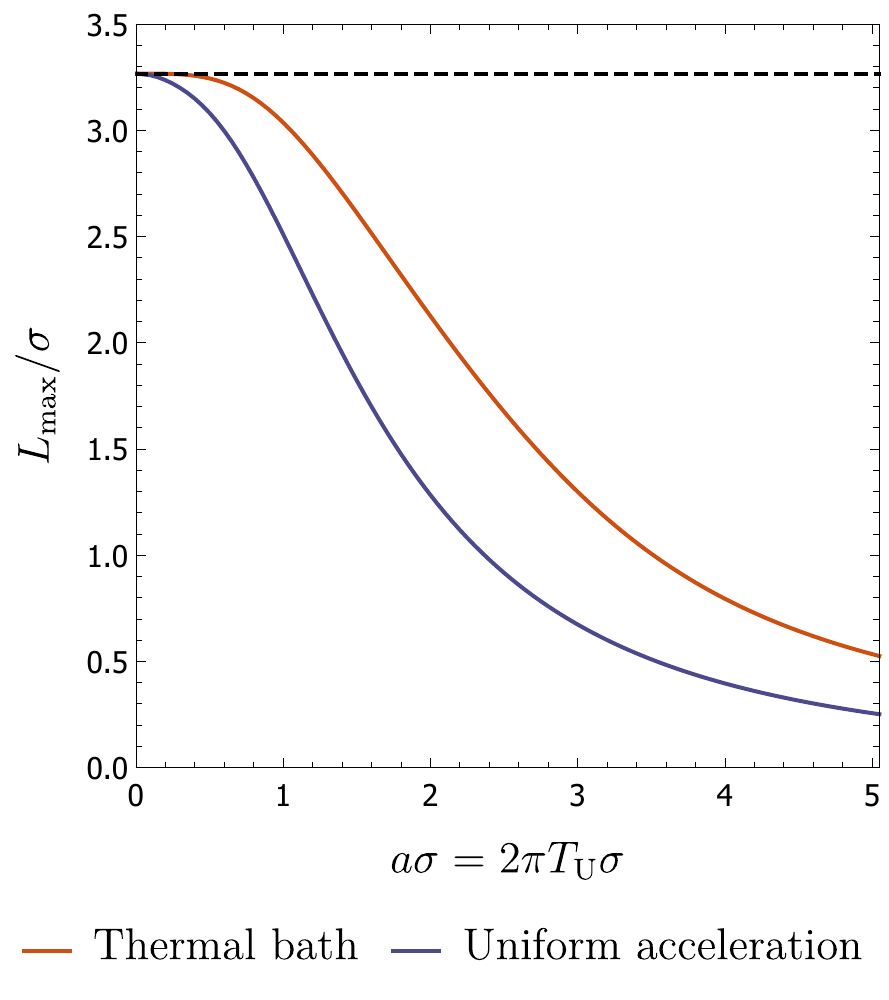}}\;
 \subfloat[$\Omega\sigma=3.00$]{\label{Lmax-d13}\includegraphics[width=0.315\linewidth]{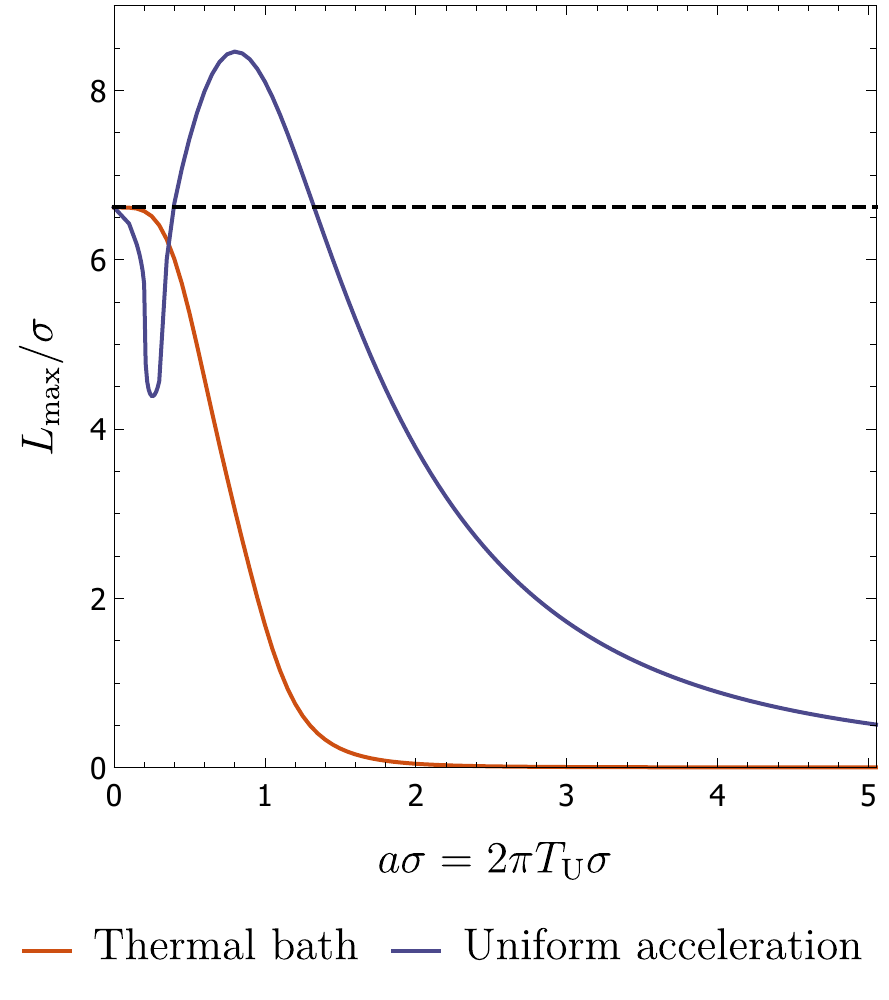}}
  \caption{The maximum separation, $L_{max}$, between
two detectors when entanglement harvesting almost does not occur is
plotted as a function of the  acceleration (or the Unruh
temperature). Here, we have set $\Omega\sigma =\{0.50, 1.00,3.00\}$.
The dashed horizon line shows the case of $a=0$, i.e., the scenario
of inertial detectors in Minkowski vacuum. }\label{Lmax-d}
\end{figure}

Now, we are in a position to analyze how two situations differ in the harvesting-achievable range of
interdetector separation.   We introduce a parameter,
$L_{max}$, to characterize the maximum harvesting-achievable
interdetector separation, beyond which entanglement harvesting
cannot occur. As shown in Fig.~\ref{Lmax-d}, for a small  energy
gap ($\Omega\sigma<1$), $L_{max}$ for both the scenario of detectors in uniform
acceleration in vacuum and  that of static ones in  a thermal bath  is a
monotonically decreasing function of  acceleration or the Unruh
temperature, and the thermal bath scenario may possess a
comparatively large harvesting-achievable range.

Notably, it
was argued in Ref.~\cite{Salton-Man:2015} that the degradation of
entanglement harvesting for two uniformly accelerated detectors is
different from that of two inertial detectors in a thermal bath.
More specifically, according to
Eqs.~(3.1) and~(3.3) and  Fig.~2 in
Ref.~\cite{Salton-Man:2015}, which are based on the saddle-point approximation, one can  infer that the static
detectors in the Minkowski thermal bath at the Unruh temperature
have a comparatively larger harvesting-achievable range than the
(parallel) accelerated ones.
However,  our numerical plots in Fig.~\ref{Lmax-d} show that this conclusion is not
universal but contingent upon a small energy gap. In fact, for a  large enough energy
gap ($\Omega\sigma\gg1$), the acceleration scenario may instead have a comparatively large $L_{max}$
[see Fig.~\ref{Lmax-d}(c)].

It is rather interesting  to note that for
$\Omega\sigma\gg1$ the $L_{max}$ curve in the acceleration case is not
an exactly decreasing function of  acceleration but exhibits some
oscillation. This reveals that the detectors in uniform acceleration can
possess a comparatively larger entanglement harvesting-achievable
interdetector separation than those in a thermal bath when the
energy gap is sufficiently  large although the static detectors enjoy a
comparatively larger harvesting-achievable separation when the
energy gap is small as was found previously. More remarkably,
acceleration can even enlarge the harvesting-achievable range of
interdetector separation as compared to the inertial
vacuum case. In contrast, the noise of a thermal bath can never
enlarge the maximum harvesting-achievable interdetector separation.
It is worth pointing out that the oscillation is an indication of the  anti-Unruh
effect again in terms of
entanglement harvesting for accelerated detectors since it means
nonmonotonicity of the harvesting-achievable separation as
acceleration varies.

\begin{figure}[!htbp]
\centering
\includegraphics[width=0.7\linewidth]{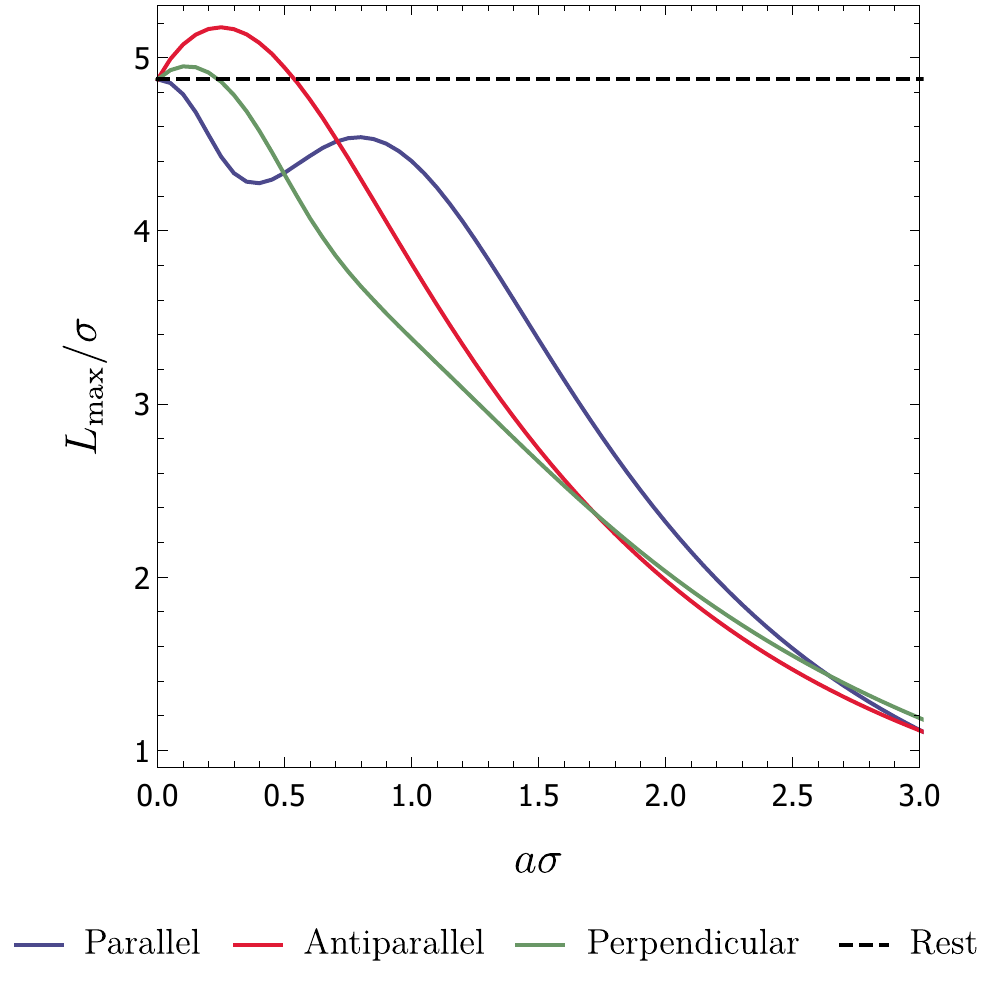}
\caption{The maximum harvesting-achievable interdetector separation, $L_{max}$, is plotted as a function of  acceleration for  parallel, antiparallel and mutually perpendicular
acceleration scenarios. Here, we have set $\Omega\sigma=2.00$, and the dashed curve denotes the case of detectors at rest. }\label{Lmaxvsa}
\end{figure}

Noteworthily,  although only the entanglement harvesting of accelerated detectors in the  parallel acceleration scenario is compared with that of inertial detectors in a thermal bath,  our conclusions remain qualitatively unchanged if  other acceleration scenarios,  e.g., those of antiparallel and mutually perpendicular
acceleration, are instead adopted. This is because, for a sufficiently large energy gap of the detectors,
acceleration  increases, in all the scenarios, the amount of harvested entanglement and
enlarges the harvesting-achievable range~\cite{Zhjl:2022},  in contrast to a previous claim~\cite{Salton-Man:2015}.   Furthermore, the harvesting-achievable range is not a monotonic function of acceleration in all the scenarios as one can see from Fig.~\ref{Lmaxvsa}, in which we plot $L_{max}$ versus $a\sigma$ for parallel, antiparallel and  mutually perpendicular
acceleration scenarios.

\section{conclusion}
Within the framework of the entanglement harvesting
protocol, we have made a detailed comparison between entanglement
harvesting in the sense of both the amount of entanglement
harvested and  the harvesting-achievable separation range  for uniformly accelerated detectors in
vacuum and static ones in a  thermal bath at the Unruh
temperature and find that
the entanglement harvesting for two uniformly accelerated detectors
is markedly different from that for two static detectors in a thermal
bath,  suggesting that equivalence between acceleration and a thermal bath in terms of the response of a single detector  is lost for the entanglement harvesting for two detectors.

Regarding the amount of entanglement harvested by detectors,
we find  that static detectors in a thermal bath always
harvest less entanglement than  those in vacuum. In
other words, no thermal noise-assisted entanglement harvesting ever
occurs. In contrast, there exists acceleration-assisted entanglement harvesting
 as long as the energy gap is large enough.  A cross-comparison shows that
static detectors  in a thermal bath can harvest more entanglement
than uniformly accelerated ones if the energy gap of the detectors is
much smaller than the detectors' Heisenberg energy, while as the
energy gap grows large enough, the accelerated detectors may  instead harvest
comparatively more entanglement.  There is a critical value of
interdetector separation, below which accelerated  detectors can acquire comparatively more entanglement. It
is interesting  to note that the critical interdetector separation  is  in
general a decreasing function of  acceleration or the Unruh
temperature, and  however it has a chance to become larger than the effective inertial regime ($\sim 1/a$) when the energy gap is large enough, signaling the phenomenon of acceleration-assisted entanglement harvesting.

With respect to the harvesting-achievable separation range,
we find that, for a small energy gap ($\Omega\sigma<1$) the thermal bath
scenario  possesses a comparatively larger harvesting-achievable
range than the uniform acceleration scenario, which is in accordance
with that obtained in Ref.~\cite{Salton-Man:2015} based on
the saddle-point approximation.  However,  for a large enough
energy gap ($\Omega\sigma\gg1$), the accelerated
detectors have, in contrast, a larger harvesting-achievable separation
range.  Moreover, acceleration can  even
enlarge the harvesting-achievable range of  interdetector
separation in comparison with the inertial vacuum case for a large
energy gap.  However,   thermal noise  can
never enlarge the harvesting-achievable range but only monotonically
 shorten the
harvesting-achievable interdetector separation  as  temperature increases.

 A notably interesting feature is that,
 although both the amount of entanglement harvested and the harvesting-achievable interdetector separation for static detectors in a thermal bath are always a decreasing function of temperature, they  are not  always so for uniformly accelerated detectors as acceleration (Unruh temperature) varies. In fact, the amount of entanglement harvested
may first increase and then decrease in the regime of small acceleration for sufficiently large energy gap, while  the harvesting-achievable interdetector separation exhibits some oscillation  in the regime of small acceleration for sufficiently large energy gap.

Finally, let us stress that  the nonmonotonicity of  the two physical quantities characterizing the entanglement harvesting phenomenon in the acceleration case (i.e., the amount of entanglement harvested and harvesting-achievable interdetector separation)  as a function of acceleration, which physically indicates that the  existence of the anti-Unruh
effect  in terms of  entanglement harvesting,
 and  acceleration-assisted enhancement of the amount of entanglement harvested and enlargement of the harvesting-achievable interdetector separation are two remarkable properties which
 distinguish the entanglement harvesting of the accelerated detectors from that of inertial detectors in a thermal bath, since the amount of entanglement harvested and  the harvesting-achievable interdetector separation are all monotonic functions of temperature and are never enhanced or enlarged by thermal noise.

%%%%%%%%%%%%%%%
 \begin{acknowledgments}
 We are grateful to Jiawei Hu for a number of helpful discussions. This work was supported in part by the NSFC under Grants  No.~12175062 and No.~12075084
 and the Research Foundation of Education Bureau of Hunan Province, China, under Grant
 No.~20B371.
\end{acknowledgments}

\def\ACP{AIP Conf. Proc.}
\def\AIHP{Ann. Inst. Henri. Poincar\'e}
\def\AJP{Amer. J. Phys.}
\def\AM{Ann. Math.}
\def\AP{Ann. Phys. (N.Y.)}
\def\APJ{Astrophys. J.}
\def\ASS{Astrophys. Space Sci.}
\def\ATMP{Adv. Theor. Math, Phys.}
\def\CJP{Can. J. Phys.}
\def\CMP{Commun. Math. Phys.}
\def\CPB{Chin. Phys. B}
\def\CPC{Chin. Phys. C}
\def\CPL{Chin. Phys. Lett.}
\def\CQG{Classcal Quantum Gravity}
\def\CTP{Commun. Theor. Phys.}
\def\EASPS{EAS Publ. Ser.}
\def\EPJC{Eur. Phys.  J. C.}
\def\EPL{Europhys. Lett.}
\def\GRG{Gen. Relativ. Gravit.}
\def\IJGMMP{Int. J. Geom. Methods Mod. Phys.}
\def\IJMPA{Int. J. Mod. Phys. A}
\def\IJMPD{Int. J. Mod. Phys. D}
\def\IJTP{Int. J. Theor. Phys.}
\def\JCAP{J. Cosmol. Astropart. Phys.}
\def\JGP{J. Geom. Phys.}
\def\JETP{J. Exp. Theor. Phys.}
\def\JHEP{J. High Energy Phys.}
\def\JMP{J. Math. Phys. (N.Y.)}
\def\JPA{J. Phys. A}
\def\JPCS{J. Phys. Conf. Ser.}
\def\JPSJ{J. Phys. Soc. Jap.}
\def\LMP{Lett. Math. Phys.}
\def\LNC{Lett. Nuovo Cim.}
\def\MPLA{Mod. Phys. Lett. A}
\def\NPB{Nucl. Phys. B}
\def\PCAM{Proc. Symp. Appl. Math.}
\def\PCPS{Proc. Cambridge Philos. Soc.}
\def\PDU{Phys. Dark Univ.}
\def\PLA{Phys. Lett. A}
\def\PLB{Phys. Lett. B}
\def\PR{Phys. Rev.}
\def\PRA{Phys. Rev. A}
\def\PRD{Phys. Rev. D}
\def\PRE{Phys. Rev. E}
\def\PRL{Phys. Rev. Lett.}
\def\PRX{Phys. Rev. X}
\def\PRSLA{Proc. Roy. Soc. Lond. A}
\def\PTP{Prog. Theor. Phys.}
\def\PRp{Phys. Rept.}
\def\RMP{Rev. Mod. Phys.}
\def\SB{Sci. Bull.}
\def\SPP{Springer Proc. Phys.}
\def\SRTU{Sci. Rep. Tohoku Univ.}
\def\ZPC{Zeit. Phys. Chem.}

\end{document}